\documentclass[noeprint,twocolumn,showpacs,amsmath,amssymb,sort&compress,floatfix,superscriptaddress,prl,aps]{revtex4-1}

\usepackage{graphicx}
\usepackage{dcolumn}
\usepackage{bm,ulem}
\usepackage{hyperref}
\usepackage[mathlines]{lineno}
\usepackage{mathtools}
\usepackage{amsmath}
\usepackage{float}
\usepackage[caption=false]{subfig}
\usepackage[english]{babel}
\usepackage{xcolor}
\usepackage{xparse}
\newcommand{\abs}[1]{\left|#1\right|}
\newcommand{\avg}[1]{\left\langle#1\right\rangle}
\renewcommand{\vec}[1]{\bm{#1}}

\renewcommand{\doibase}[1]{https://dx.doi.org/\ifdefempty{#1}{}{#1}}

\usepackage{etoolbox}
\let\mybibitem\bibitem
\renewcommand{\bibitem}[1]{%
\ifstrequal{#1}{SM}{\color{black}\mybibitem{#1}}
  {\ifstrequal{#1}{Eisenhoffer2012}{\color{black}\mybibitem{#1}}
    {\ifstrequal{#1}{Fily2014}{\color{black}\mybibitem{#1}}
      {\ifstrequal{#1}{Staple2010}{\color{black}\mybibitem{#1}}
        {\ifstrequal{#1}{Etournay2015}{\color{black}\mybibitem{#1}}
          {\ifstrequal{#1}{Nagai2001}{\color{black}\mybibitem{#1}}
            {\ifstrequal{#1}{Li2014b}{\color{black}\mybibitem{#1}}
              {\ifstrequal{#1}{Saw2017}{\color{black}\mybibitem{#1}}
              {\color{black}\mybibitem{#1}}
              }
            }
          }    
        }
      }
    }
  }%
}

\makeatletter
\renewcommand{\fnum@figure}{FIG. \thefigure}
\makeatother

\newcommand{\bl}[1]{{\color{black}#1}}

\newcommand{\ucsb}{Department of Physics, University of California Santa Barbara, Santa Barbara, CA 93106, USA}
\newcommand{\ue}{SUPA, School of Physics and Astronomy, University of Edinburgh, Peter Guthrie Tait Road, Edinburgh EH9 3FD, United Kingdom}

\begin{document}

\title{Solid-Liquid Transition of Deformable and Overlapping Active Particles}

\author{Benjamin Loewe}
\affiliation{\ucsb}
\author{Michael Chiang}
\affiliation{\ue}
\author{Davide Marenduzzo}
\affiliation{\ue}
\author{M. Cristina Marchetti}
\affiliation{\ucsb}

\begin{abstract}
Experiments and theory have shown that cell monolayers and epithelial tissues exhibit solid-liquid and glass-liquid transitions. These transitions are biologically relevant to our understanding of embryonic development, wound healing, and cancer. \bl{Current models of confluent epithelia  have focused on the role of cell shape, with less attention paid to  cell extrusion, which is key for maintaining homeostasis in biological tissue.} Here, we use a multi-phase field model to study the solid-liquid transition in a confluent monolayer of deformable cells. \bl{Cell overlap is allowed and provides a way for modeling the precursor for extrusion}. When cells overlap rather than deform, we find that the melting transition changes from continuous to \bl{first-order-like}, and that there is an intermittent regime close to the transition, where solid and liquid states alternate over time. By studying the dynamics of $5$- and $7$-fold disclinations in the hexagonal lattice formed by the cell centers, we observe that these correlate with spatial fluctuations in the cellular overlap, and that cell extrusion tends to initiate near $5$-fold disclinations.

\end{abstract}

\maketitle

Understanding the dynamics and collective behavior of cells in dense tissues is an important goal of biophysics, with relevance to a number of developmental processes, such as embryogenesis~\cite{Chuai2012}, wound healing~\cite{Poujade2007}, and cancer~\cite{Haeger2014}. For example, the epithelial-mesenchymal transition  can be viewed as a solid-liquid transition occurring {\it in vivo} \cite{Thiery2002, Thompson2005, Mitchel2019}, where cells become more motile and less adhesive: this transition has been reported to play a role in tissue repair, inflammation, and \bl{tumor} progression~\cite{EMT,Friedl2009,Haeger2014}. Experimental studies have also shown that epithelial cells can undergo an \bl{unjamming} transition between a glassy phase where their dynamics is slow to a fluid phase with large-scale collective motion both {\it in vitro}~\cite{Angelini2011,Nnetu2012,Park2015,Garcia2015, Malinverno2017, Atia2018} and {\it in vivo} \cite{Mongera2018, Atia2018}.
 
From a theoretical point of view, an appealing model of a dense tissue is provided by a two-dimensional (2D) confluent cell monolayer (i.e., a space-filling cell monolayer with packing fraction equal to unity). This system can be studied by the cellular Potts model~\cite{Graner1992}, the vertex~\bl{\cite{Nagai2001,Staple2010, Fletcher2014,Bi2015}} and Voronoi~\bl{\cite{Li2014b, Bi2016}} models, and their variants~\cite{Teomy2018c,Yan2019}. Such frameworks have recently been used to study the melting transition in monolayers of passive~\cite{Bi2015,Li2018,Durand2019} and active/self-motile cells~\cite{Bi2016,Giavazzi2018,Barton2017,Chiang2016}. Cell motility and deformability distinguish this problem from the 2D melting of crystals of hard or soft disks~\cite{Engel2013,Kapfer2015,Hajibabaei2019,Digregorio2019}, which proceeds either via a discontinuous transition~\cite{Saito1982,Chui1983}, or through an intermediate hexatic phase and the unbinding of topological defects~\cite{Kosterlitz1973,Halperin1978, Nelson1978,Young1979,Dash1999,Gasser2009}. 

Existing studies \bl{of vertex and Voronoi models} of confluent active monolayers suggest that a continuous solid-liquid (or glass-liquid) transition can be observed upon increasing cell motility~\bl{\cite{Bi2016, Barton2017}}. While useful in providing quantitative predictions, \bl{this work has mainly focused on the role of cell intercalation (T1 transitions) in controlling tissue rigidity and less on the role of cell extrusion that in these strictly $2D$ models may be described by cell removal (T2 transitions) ~\cite{Staple2010, Etournay2015}. 
In many situations, however, cell extrusion is driven by cell crowding and overlap, as commonly seen in confluent epithelia ~\cite{Eisenhoffer2012}.} 
Cell overlap  also occurs during early embryogenesis as \bl{an epithelial} monolayer is converted into a multi-layered epithelium following a tightly coordinated stratification program~\cite{Koster2007}. 
\bl{Here we consider a model that explicitly allows for cell overlap, interpreted as a precursor for cell extrusion, to examine its role on the solid-liquid transition of a confluent tissue.}

\bl{To incorporate} both particle deformation and overlap \bl{we use a} multi-phase field model \cite{Nonomura2012,Palmieri2015,Foglino2017,Mueller2019} to study melting of a confluent \bl{layer of motile} deformable particles. \bl{The behavior of our system is controlled by the} trade-off between deformability and overlap: the less deformable a particle is, the more it overlaps with its neighbors. \bl{At high deformability we find a continuous solid-liquid transition with increasing cell motility. The transition becomes first-order-like at low deformability when cells  overlap, with
an intermediate intermittent state, where the system as a whole alternates between solid and liquid states.} Finally, we observe a strong correlation between unbound structural defects (corresponding to $5$- and $7$-fold disclinations in the hexagonal lattice formed by the cell centers) generated upon melting and local fluctuations in cell overlap. Specifically, we find that cellular extrusion is \bl{favored at $5$-fold disclinations.} 

Our model may also serve as a bridge between particle-based and confluent models. 
\bl{Upon decreasing cell deformability, the system transitions from deformable particles that tessellate their domain without overlap, similar to vertex models, to  almost-circular overlapping disks.} \bl{The connection with these two limiting cases is, however, only qualitative. At high deformability,  anisotropy of cell shape is strongly correlated with fluidity \bl{(Figs.~S5 and S6)}, but, unlike vertex models, it does not provide an order parameter for the liquid state. Conversely, at low deformability, overlap in our model is much higher than that allowed in systems of soft disks.}  
\begin{figure}[t]
  \includegraphics[width=\columnwidth]{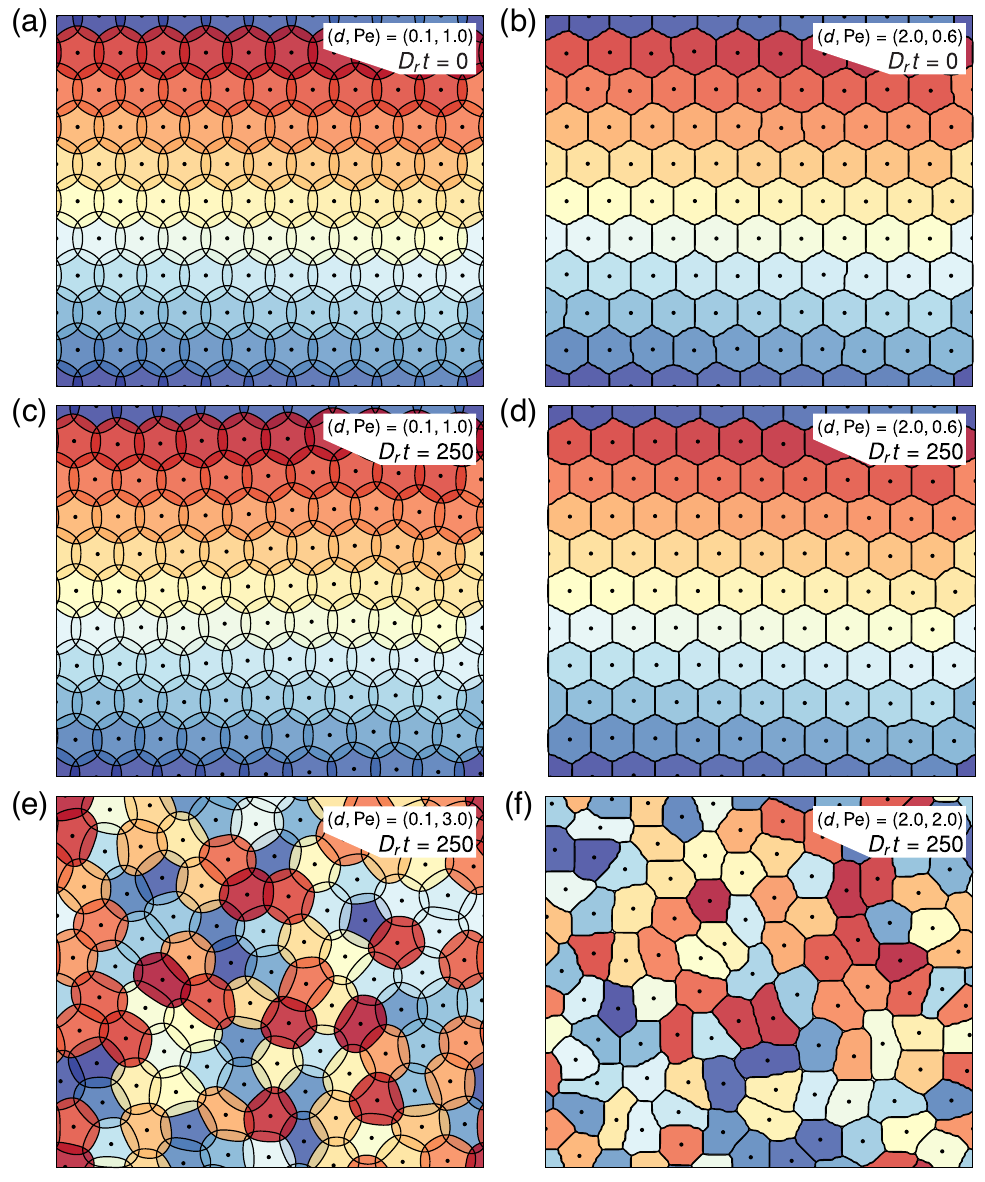}
  \caption{Simulation snapshots of the stationary state for different deformability (quantified by $d$) and motility (quantified by $\text{Pe}$). The contours of the cells correspond to the level $\{\phi_i = 1\}_{i=1}^{N}$, while the coloring corresponds to cell-index at $t = 0$ (for visualizing cell rearrangements). (a)-(b) The initial condition of the monolayer at (a) $d = 0.1$ and (b) $d = 2.0$. Note that cells overlap at low $d$, whereas at high $d$ cells deform rather than overlap. (c)-(f) Snapshots of the system at $D_rt = 250$. The system remains in a crystal-like state at low motility ((c) and (d)). At sufficiently high motility, the system melts and cells exchange neighbors ((e) and (f)). See also Suppl. Movies 1-4.}
  \label{Fig1}
\end{figure}

Our multi-phase field model contains $N$ scalar fields, $\{\phi_i(\vec{r})\}_{i=1}^N$, each representing a different cell. The equilibrium configuration of the cell layer is determined by the minimization of the following free energy~\cite{Palmieri2015,Mueller2019}:
\begin{equation}
\label{Free_Energy}
\begin{split}
  &\mathcal{F} = \sum_{i=1}^N\Bigg[\int d^2\vec{r}\left(\frac{\alpha}{4}\phi_i^2(\phi_i-\phi_{0})^2+\frac{K}{2}(\vec{\nabla}\phi_i)^2\right)\\
  &+\lambda\left(1-\int d^2\vec{r}\,\frac{\phi_i^2}{\pi R^2 \phi_0^2}\right)^2+\varepsilon\sum_{i<j=1}^N\int d^2\vec{r}\,\phi_i^2 \phi_j^2\Bigg]. 
\end{split}
\end{equation}
\bl{The first three terms} determine the shape of the cells. The first term sets $\phi_0$ and zero as the preferred values of the field inside and outside the cell, respectively. The second term penalizes spatial variations of $\phi$. Together, they determine the physical properties of the cell boundary, such as the interfacial thickness, which we define as $\xi = \sqrt{2K/\alpha}$, and surface tension  $\sigma = \sqrt{8K\alpha/9}$ \cite{Pagonabarraga2002}. The third term is a soft constraint that sets the preferred area of the cell to that of a circle with target radius $R$. Finally, the fourth term models the steric repulsion between cells by energetically penalizing cell overlap. 

To model the dynamics of self-motile active cells, we assume simple relaxational and overdamped dynamics,
\begin{equation}\label{phieq}
\frac{\partial \phi_i}{\partial t} +\vec{v}_i\cdot\vec{\nabla}\phi_i = -\frac{1}{\gamma}\frac{\delta\mathcal{F}}{\delta\phi_i}, 
\end{equation}
where $\gamma$ is a friction \bl{coefficient} and we have included an advection term that propels the cells with velocity $\vec{v}_i$ \bl{, see~\cite{SM}}. All cells have the same  propulsion speed $v_0$, while their direction of motion $\theta_i$ is controlled by rotational noise with diffusivity $D_r$, 
\begin{equation}\label{rotationaldiffusion}
    d\theta_i(t) = \sqrt{2 D_r}\,dW_i(t), 
\end{equation}
where $W_i$ is a Wiener process. Cell motility is quantified by the P\'eclet number $\text{Pe} \equiv (v_0/D_r)/R$, which is the ratio between the cells' persistence length and their target radius. These equations are a generalization of the active Brownian particle model~\cite{Romanczuk2012,Fily2012} to a system of deformable cells.

Our model allows cells to both deform and overlap. In general, these are competing effects: deformation is energetically \bl{penalized by surface tension, while overlap is penalized by repulsion}. \bl{We quantify cell deformability }through the dimensionless ratio $d\equiv\varepsilon/\alpha$. When $d\ll 1$,  cells tend to acquire a circular shape and overlap with their neighbors (Fig.~\ref{Fig1}a,c,e). Conversely,  when $d\gg 1$, cells  change their shape to match with their neighbors and minimize overlap (Fig.~\ref{Fig1}b,d,f). 

We first examine the role of deformability and motility on the solid-liquid transition at confluence.  To this end, we employ a finite difference method to solve numerically Eqs.~(\ref{phieq}) and (\ref{rotationaldiffusion}) for $N = 36$ and $100$ cells \bl{in a rectangular box of aspect ratio that accommodates an undeformed  hexagonal cell lattice}, with periodic boundary conditions. 
Choosing $R$ as unit of length and $D_r^{-1}$ as  unit of time, \bl{we use}  $\delta x=1/12$ and $\Delta t=5\times10^{-5}$ \bl{as our simulation lattice unit and timestep, respectively}.
\bl{We tune deformability by varying $\alpha$ and $K$ such that $\xi$ is constant.}  We initiate the cells in a hexagonal lattice \bl{ with $\lambda \geq 3000\,K$} and allow the system to achieve confluence by setting the cell target area $\pi R^2$ to be larger than the area available to each cell.  Further simulation details and the list of parameters are given in the supplemental material.

\begin{figure}
  \includegraphics[width=\columnwidth]{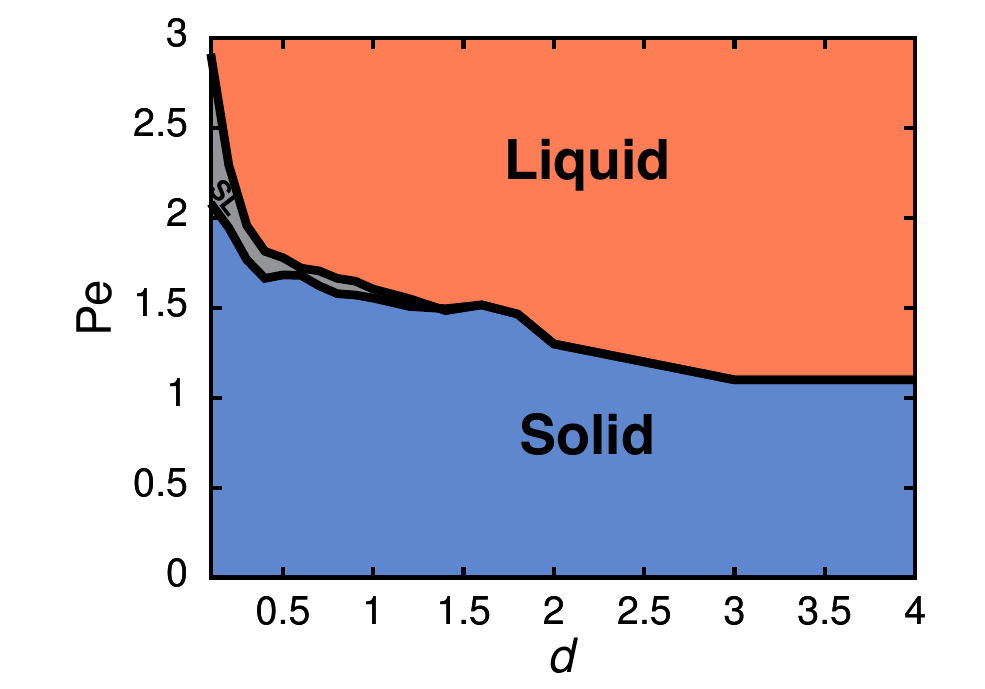}
  \caption{Phase diagram of melting in our confluent system. The transition lines separating the solid, intermittent (labeled as SL), and liquid phases are interpolation of the boundaries identified based on the system's diffusivity $\overline{D}_{\text{eff}}$, its global bond-orientational order $\abs{\Psi_6}$, and the fraction of structural defects $\avg{\abs{\Delta N_{\text{nn}}}}$ in the system.}
  \label{Fig2}
\end{figure}

To quantify the melting transition, we compute both dynamical and structural observables~\bl{\cite{SM}.} Dynamical arrest is quantified  through an effective diffusivity $\overline{D}_{\text{eff}}$~\cite{Bi2016, Giavazzi2018} obtained from the long-time behavior of the mean square displacement $\text{MSD}(t)$ of individual cells as \bl{$\overline{D}_{\text{eff}} = \lim_{t\rightarrow\infty} \text{MSD}(t)/(4D_0t)$,} with $D_0 = v_0^2/(2 D_r)$ the diffusivity of an isolated cell. As structural observables, we measure the global bond-orientational order parameter $\abs{\Psi_6}$ and the structure factor $S(\vec{q})$. Choosing $\overline{D}_{\text{eff}}>0.0005$ as the threshold for a liquid state, the transition lines obtained from the dynamical and structural measurements  coincide. The phase diagram displayed in Fig.~\ref{Fig2} shows that both deformability and motility facilitate melting. We also find a  region of intermittence at low deformability, discussed further below. The width of the plateau in the $\text{MSD}$ at intermediate times shrinks with increasing deformability, suggesting that  deformability facilitates melting by allowing particles to squeeze more easily through the cages provided by their neighbors.

One of our key results is that the nature of the transition is different at low and high deformability. This can be appreciated by analyzing the standard error of $\abs{\Psi_6}$ across the parameter space $(d,\text{Pe})$, which shows that there is an intermediate $\text{Pe}$ range at $d<1$ for which this quantity is large.  
Intriguingly, $d<1$ is precisely the region in parameter space where the overlap between cells becomes appreciable, implying that the character of monolayer melting depends on whether the rearrangement of particles occurs by cells squeezing past their neighbors \bl{ by deforming ($d>1$)} or crawling over them \bl{by overlapping ($d<1$)}. 

\begin{figure}[ht]
\includegraphics[width=0.5\textwidth]{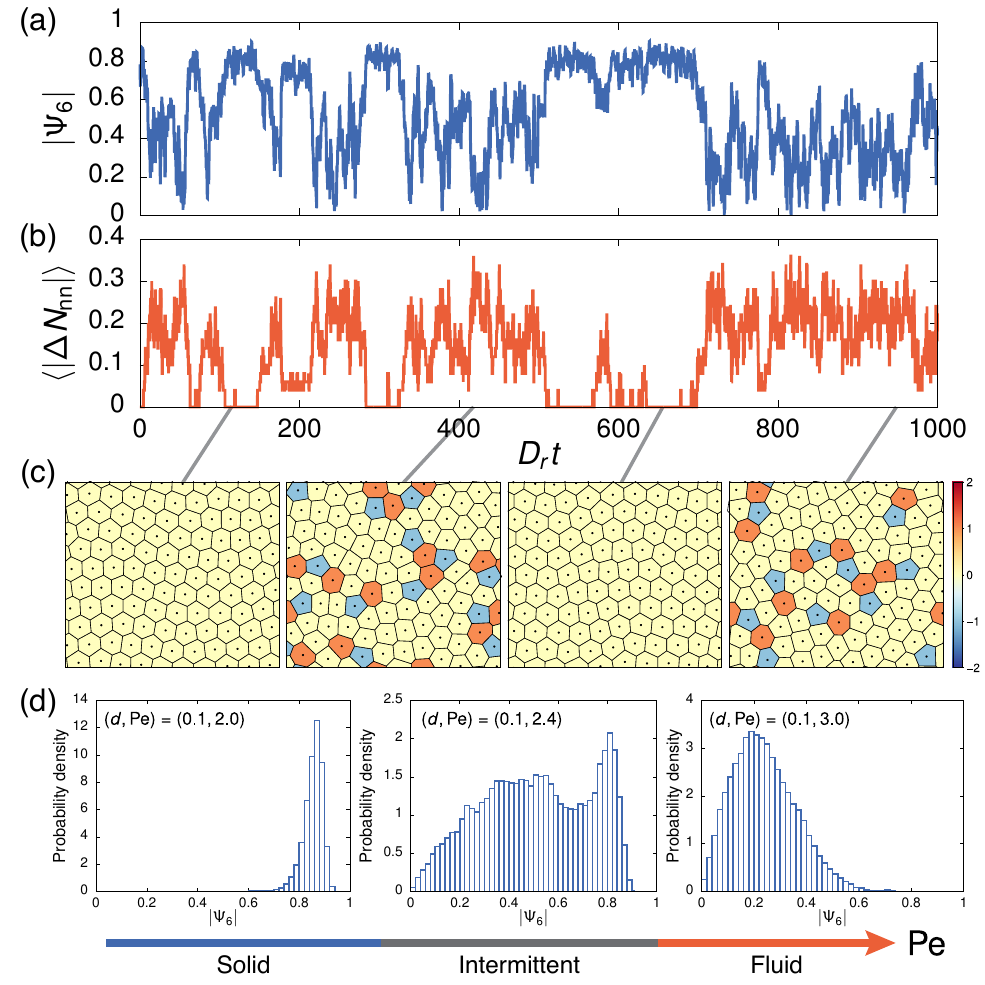}
\caption{\bl{Intermittence at low deformability. (a)-(c) Intermittent behavior at $(d,\text{Pe})=(0.1,2.4)$.} (a) Time series of $\abs{\Psi_6}$. (b) Time series of the fraction of defects $\avg{\abs{\Delta N_{\text{nn}}}}$. (c) Voronoi \bl{tessellation} of the system depicting instantaneous configurations in the solid and liquid states, and highlighting the presence of topological defects in the latter. The color scheme denotes the topological charge -- i.e., the deviation from a coordination number of six. (d) PDFs of $\abs{\Psi_6}$ at $d=0.1$ with increasing motility. As the system goes through the intermittent regime, the PDF goes from unimodal (solid), to bimodal (intermittent), and back to unimodal (liquid). }
\label{Fig3}
\end{figure}

To determine the nature of the intermediate regime found at $d<1$, we analyze the corresponding time series of $\abs{\Psi_6}$ (Fig.~\ref{Fig3}a). The time series shows clear evidence of an intermittent behavior, where the system jumps between two distinct states with different mean values of $\abs{\Psi_6}$ (see also Suppl. Movie 5). The two states are also apparent from the bimodal character of the $\abs{\Psi_6}$ probability density function (PDF; Fig.~\ref{Fig3}d). Since $\abs{\Psi_6}$ correlates with the melting transition and our solid state is close to a hexagonal crystal, we can associate $\abs{\Psi_6} \simeq 1$ to a solid state, and values of $\abs{\Psi_6}$ \bl{close to} or below $0.5$ to a liquid state. Moreover, the values of $\abs{\Psi_6}$ in the solid and liquid regimes fluctuate around well-defined means, and hence exhibit unimodal PDFs (albeit with different widths), so that bimodality in the PDF signals \bl{intermittence}. We also identify \bl{intermittence} by computing the fraction of defects in the system $\avg{\abs{\Delta N_{\text{nn}}}}$ (Fig.~\ref{Fig3}b,c), i.e., the fraction of the total number of cells with a coordination number other than six. The time series for $\avg{\abs{\Delta N_{\text{nn}}}}$ shows that defects appear when the monolayer is in the liquid state. \bl{In addition, and in line with Ref.~\cite{Digregorio2019}, we observe that defects in the intermittent phase tend to form grain boundaries and percolate the system (Figs.~3c and S11).}

We locate the intermittent region in the phase diagram (Fig.~\ref{Fig2}) via two separate methods. First, given that there are large \bl{fluctuations} in $\abs{\Psi_6}$ in this region, we identify states to be intermittent if both the standard error of $\abs{\Psi_6}$ and $\overline{D}_{\text{eff}}$ are above $0.0005$. Second, we binarize the time series of $\avg{\abs{\Delta N_{\text{nn}}}}$ and map each time point to either zero (solid) or unity (liquid). For a time series to be intermittent, we require a minimum of two jumps between the states, and a large enough fraction of time spent in either state. Both methods converge and pinpoint a similar parameter region to be intermittent. Further, this region shrinks with increasing $N$ \bl{(see Figs.~S8 and S10). This} suggests that \bl{intermittence} arises because the solid-liquid transition is first-order-like at low deformability, so that coexistence between the two phases is expected at criticality. \bl{The first order character is also supported by  finite size scaling of $|\Psi_6|  = N^{\zeta} f(p N^{\nu})$ at low $d$, where $p = \text{Pe}/\text{Pe}^*-1$ and $\zeta=-0.044(3)$, computed for systems up to 900 cells (see Fig.~S12). }

As anticipated, and clear from the phase diagram, the intermittent phase is only present at low deformability, when cells overlap. A possible mechanism through which cell overlap might affect the nature of the transition is the following. When cells are highly deformable and do not overlap with their neighbors, they can escape the local cage in which they are trapped by squeezing through  their neighbors. These cage escapes lead to neighbor exchanges, hence to fluidification. On the other hand, if cells are not deformable but can overlap, moving a cell is similar to inserting or moving a coin on a substrate crowded with other coins (as in a ``coin-pusher'' arcade game). In this case, motion of the coin can either result in simple coin overlap/layering and no motion, or in the collective motion of a raft of coins. The coexistence of different scenarios (overlap or collective motion) may underlie the onset of \bl{intermittence} in our simulations, and the first-order-like nature of the solid-liquid transition in the low deformability regime. \bl{We note a first-order-like glass-to-liquid transition has also been found in systems of active soft disks~\cite{Fily2014}}.

Finally, we analyze the relation between defects in the bond-orientational order and cell overlap. Experiments with monolayers of progenitor stem cells \cite{Kawaguchi2017} have shown that these systems can be viewed as active nematics, and that topological defects in the nematic order correlate with the location of cell  extrusion and death. Similar behavior has been obtained in MDCK (Madin Darby Canine Kidney) cells~\cite{Saw2017, Mueller2019}. On the other hand, nematic order is often not readily apparent in epithelia, where cells are typically not elongated,  and extrusion is presumably associated with  high local overlap of a cell with its neighbors~\cite{Eisenhoffer2012}. Our work offers an alternative interpretation that  correlates cell extrusion not with defects in nematic order, but with cell overlap and associated structural defects in cell packing. 

\begin{figure}[t]
\includegraphics[width=\columnwidth]{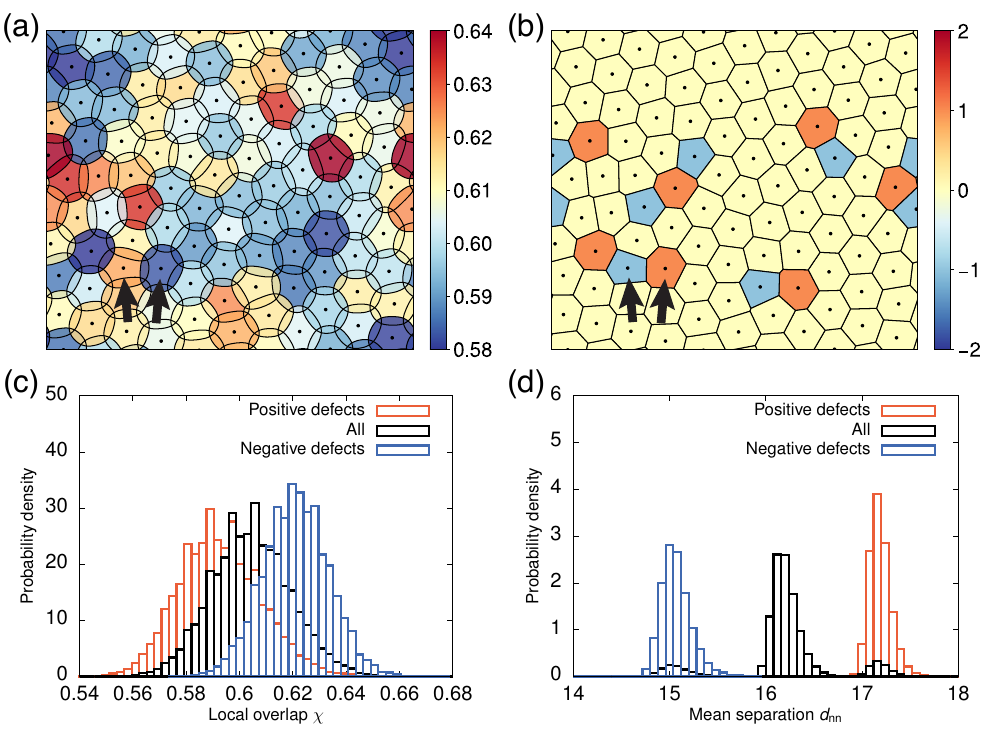}
\caption{Correlation between local overlap and defects at $(d, \text{Pe})=(0.1,2.4)$. (a) \bl{Simulation} snapshot. Color denotes the degree of local overlap $\chi$. (b) Voronoi-\bl{tessellation} of (a), colored by the topological charge of each cell (see Fig.~\ref{Fig3}c). Black arrows point to the same two cells with defects in (a) and (b) to highlight that cells with $5$-fold disclinations have higher local overlap than those with $7$-fold disclinations. (c) PDFs of the local \bl{overlap for} different defect types. \bl{(d) PDFs of the mean separation $d_{\text{nn}}$ of a cell from its nearest neighbors for different defect types.}}
\label{Fig4}
\end{figure}

Defects in the hexagonal lattice  formed by the cell centers in the ordered solid state are  $5$-fold and $7$-fold disclinations and correspond  to pentagonal and heptagonal cells, respectively, in the associated Voronoi \bl{tessellation} ~\cite{Bowick2009}. They are readily identified in the cell packing, as shown in \bl{Fig.~\ref{Fig4}a,b}. We define the local overlap of the $i$th cell as $\chi_i(\vec{r})\equiv\sum_{j=1}^NH(\phi_i(\vec{r})-1)H(\phi_j(\vec{r})-1)$, with $H$ the Heaviside function. We then search for correlations between defects and overlap by recording both overlap and coordination number for each cell, and constructing the PDFs for the local overlap for pentagonal/heptagonal  cells, as well as for the entire cell population \bl{(Fig.~\ref{Fig4}c)}. The PDFs show that pentagons experience, on average, \bl{more overlap with respect to other cells.} \bl{This can be understood by noting that, while all cells have approximately the same area (Fig. 4a), 5-fold coordinated particles have a smaller mean distance to their neighbors (Fig. 4d). Hence cell overlap is largest at 5-fold defects, suggesting that these may be likely loci of cell extrusion, which is known that can be triggered by cell crowding. Our results suggest that cell extrusions in cell monolayers  are likely to occur in the intermittent regime or near the solid-liquid transition, and may originate near 5-fold coordinated}  cells.

In summary, we have used a multi-phase field model to explore the effect of overlap and motility on the solid-liquid transition in confluent monolayers of active deformable cells. \bl{Melting is} triggered by increasing motility and/or deformability, which \bl{promotes fluidification} by allowing cells to squeeze past their neighbors. We have shown that overlap strongly affects the nature of the melting transition in the monolayer. Specifically, when cells overlap rather than deform, the solid-liquid transition changes from continuous to first-order-like, and it is accompanied by an intermediate intermittent regime in which the monolayer alternates between  solid and liquid states. This intermittent phase could be relevant to morphological processes that require periodic fluidization to restructure the tissue. We have also found a correlation between the location of topological defects in cell packing and \bl{fluctuations} in local cell overlap, which suggests that cellular extrusion could be linked to the presence of these defects. Extrusion is an important process in epithelial tissues required for proper biological functioning. While it is normally thought that extrusion is determined by biochemical signaling, recent experiments have suggested a correlation between extrusion and topological defects in the orientational order of elongated or spindle-like cells. Here we suggest an \bl{alternative}, possibly more general, correlation between extrusion and topological defects in the structure of cell packings that applies even when cells are not elongated. 
 
From a theoretical point of view, it would be of interest to ask whether our active monolayers of deformable cells also exhibit a hexatic phase, which has been recently found in high-density suspensions of active Brownian particles~\cite{Digregorio2019,Digregorio2018}. Addressing this question will require simulations of much larger systems.

\paragraph*{Acknowledgements --} \bl{This work was supported by the National Science Foundation grants DMR-1609208 (B.L. and M.C.M) and PHY-1748958 (KITP). B.L. would like to acknowledge the hospitality of KITP, where some of this work was done.} M.C. acknowledges the Carnegie Trust for the Universities of Scotland for PhD studentship funding. Use was made of computational facilities purchased with funds from the National Science Foundation (CNS-1725797) and administered by the Center for Scientific Computing (CSC). The CSC is supported by the California NanoSystems Institute and the Materials Research Science and Engineering Center (MRSEC; NSF DMR 1720256) at UC Santa Barbara.

B.L. and M.C. contributed equally to this work.


\begin{thebibliography}{57}%
\makeatletter
\providecommand \@ifxundefined [1]{%
 \@ifx{#1\undefined}
}%
\providecommand \@ifnum [1]{%
 \ifnum #1\expandafter \@firstoftwo
 \else \expandafter \@secondoftwo
 \fi
}%
\providecommand \@ifx [1]{%
 \ifx #1\expandafter \@firstoftwo
 \else \expandafter \@secondoftwo
 \fi
}%
\providecommand \natexlab [1]{#1}%
\providecommand \enquote  [1]{``#1''}%
\providecommand \bibnamefont  [1]{#1}%
\providecommand \bibfnamefont [1]{#1}%
\providecommand \citenamefont [1]{#1}%
\providecommand \href@noop [0]{\@secondoftwo}%
\providecommand \href [0]{\begingroup \@sanitize@url \@href}%
\providecommand \@href[1]{\@@startlink{#1}\@@href}%
\providecommand \@@href[1]{\endgroup#1\@@endlink}%
\providecommand \@sanitize@url [0]{\catcode `\\12\catcode `\$12\catcode
  `\&12\catcode `\#12\catcode `\^12\catcode `\_12\catcode `\%12\relax}%
\providecommand \@@startlink[1]{}%
\providecommand \@@endlink[0]{}%
\providecommand \url  [0]{\begingroup\@sanitize@url \@url }%
\providecommand \@url [1]{\endgroup\@href {#1}{\urlprefix }}%
\providecommand \urlprefix  [0]{URL }%
\providecommand \Eprint [0]{\href }%
\providecommand \doibase [0]{http://dx.doi.org/}%
\providecommand \selectlanguage [0]{\@gobble}%
\providecommand \bibinfo  [0]{\@secondoftwo}%
\providecommand \bibfield  [0]{\@secondoftwo}%
\providecommand \translation [1]{[#1]}%
\providecommand \BibitemOpen [0]{}%
\providecommand \bibitemStop [0]{}%
\providecommand \bibitemNoStop [0]{.\EOS\space}%
\providecommand \EOS [0]{\spacefactor3000\relax}%
\providecommand \BibitemShut  [1]{\csname bibitem#1\endcsname}%
\let\auto@bib@innerbib\@empty
\bibitem{Chuai2012}%
  \BibitemOpen
  \bibfield  {author} {\bibinfo {author} {\bibfnamefont {M.}~\bibnamefont
  {Chuai}}, \bibinfo {author} {\bibfnamefont {D.}~\bibnamefont {Hughes}}, \
  and\ \bibinfo {author} {\bibfnamefont {C.}~\bibnamefont {J~Weijer}},\ }\href
  {\doibase 10.2174/138920212800793357} {\bibfield  {journal} {\bibinfo
  {journal} {Curr. Genomics}\ }\textbf {\bibinfo {volume} {13}},\ \bibinfo
  {pages} {267} (\bibinfo {year} {2012})}\BibitemShut {NoStop}%
\bibitem{Poujade2007}%
  \BibitemOpen
  \bibfield  {author} {\bibinfo {author} {\bibfnamefont {M.}~\bibnamefont
  {Poujade}}, \bibinfo {author} {\bibfnamefont {E.}~\bibnamefont
  {Grasland-Mongrain}}, \bibinfo {author} {\bibfnamefont {A.}~\bibnamefont
  {Hertzog}}, \bibinfo {author} {\bibfnamefont {J.}~\bibnamefont {Jouanneau}},
  \bibinfo {author} {\bibfnamefont {P.}~\bibnamefont {Chavrier}}, \bibinfo
  {author} {\bibfnamefont {B.}~\bibnamefont {Ladoux}}, \bibinfo {author}
  {\bibfnamefont {A.}~\bibnamefont {Buguin}}, \ and\ \bibinfo {author}
  {\bibfnamefont {P.}~\bibnamefont {Silberzan}},\ }\href {\doibase
  10.1073/pnas.0705062104} {\bibfield  {journal} {\bibinfo  {journal} {Proc.
  Natl. Acad. Sci. USA}\ }\textbf {\bibinfo {volume} {104}},\ \bibinfo {pages}
  {15988} (\bibinfo {year} {2007})}\BibitemShut {NoStop}%
\bibitem{Haeger2014}%
  \BibitemOpen
  \bibfield  {author} {\bibinfo {author} {\bibfnamefont {A.}~\bibnamefont
  {Haeger}}, \bibinfo {author} {\bibfnamefont {M.}~\bibnamefont {Krause}},
  \bibinfo {author} {\bibfnamefont {K.}~\bibnamefont {Wolf}}, \ and\ \bibinfo
  {author} {\bibfnamefont {P.}~\bibnamefont {Friedl}},\ }\href {\doibase
  https://doi.org/10.1016/j.bbagen.2014.03.020} {\bibfield  {journal} {\bibinfo
   {journal} {Biochim Biophys Acta}\ }\textbf {\bibinfo {volume} {1840}},\
  \bibinfo {pages} {2386} (\bibinfo {year} {2014})}\BibitemShut {NoStop}%
\bibitem{Thiery2002}%
  \BibitemOpen
  \bibfield  {author} {\bibinfo {author} {\bibfnamefont {J.~P.}\ \bibnamefont
  {Thiery}},\ }\href {\doibase 10.1038/nrc822} {\bibfield  {journal} {\bibinfo
  {journal} {Nat. Rev. Cancer}\ }\textbf {\bibinfo {volume} {2}},\ \bibinfo
  {pages} {442} (\bibinfo {year} {2002})}\BibitemShut {NoStop}%
\bibitem{Thompson2005}%
  \BibitemOpen
  \bibfield  {author} {\bibinfo {author} {\bibfnamefont {E.~W.}\ \bibnamefont
  {Thompson}}\ and\ \bibinfo {author} {\bibfnamefont {D.~F.}\ \bibnamefont
  {Newgreen}},\ }\href {\doibase 10.1158/0008-5472.CAN-05-0616} {\bibfield
  {journal} {\bibinfo  {journal} {Cancer Res.}\ }\textbf {\bibinfo {volume}
  {65}},\ \bibinfo {pages} {5991} (\bibinfo {year} {2005})}\BibitemShut
  {NoStop}%
\bibitem{Mitchel2019}%
  \BibitemOpen
  \bibfield  {author} {\bibinfo {author} {\bibfnamefont {J.~A.}\ \bibnamefont
  {Mitchel}}, \bibinfo {author} {\bibfnamefont {A.}~\bibnamefont {Das}},
  \bibinfo {author} {\bibfnamefont {M.~J.}\ \bibnamefont {O'Sullivan}},
  \bibinfo {author} {\bibfnamefont {I.~T.}\ \bibnamefont {Stancil}}, \bibinfo
  {author} {\bibfnamefont {S.~J.}\ \bibnamefont {DeCamp}}, \bibinfo {author}
  {\bibfnamefont {S.}~\bibnamefont {Koehler}}, \bibinfo {author} {\bibfnamefont
  {J.~P.}\ \bibnamefont {Butler}}, \bibinfo {author} {\bibfnamefont {J.~J.}\
  \bibnamefont {Fredberg}}, \bibinfo {author} {\bibfnamefont {M.~A.}\
  \bibnamefont {Nieto}}, \bibinfo {author} {\bibfnamefont {D.}~\bibnamefont
  {Bi}}, \ and\ \bibinfo {author} {\bibfnamefont {J.-A.}\ \bibnamefont
  {Park}},\ }\href {\doibase 10.1101/665018} {\bibfield  {journal} {\bibinfo
  {journal} {bioRxiv}\ ,\ \bibinfo {pages} {665018}} (\bibinfo {year}
  {2019})}\BibitemShut {NoStop}%
\bibitem{EMT}%
  \BibitemOpen
  \bibfield  {author} {\bibinfo {author} {\bibfnamefont {J.~P.}\ \bibnamefont
  {Thiery}}, \bibinfo {author} {\bibfnamefont {H.}~\bibnamefont {Acloque}},
  \bibinfo {author} {\bibfnamefont {R.~Y.}\ \bibnamefont {Huang}}, \ and\
  \bibinfo {author} {\bibfnamefont {M.~A.}\ \bibnamefont {Nieto}},\ }\href
  {\doibase 10.1016/j.cell.2009.11.007} {\bibfield  {journal} {\bibinfo
  {journal} {Cell}\ }\textbf {\bibinfo {volume} {139}},\ \bibinfo {pages} {871}
  (\bibinfo {year} {2009})}\BibitemShut {NoStop}%
\bibitem{Friedl2009}%
  \BibitemOpen
  \bibfield  {author} {\bibinfo {author} {\bibfnamefont {P.}~\bibnamefont
  {Friedl}}\ and\ \bibinfo {author} {\bibfnamefont {D.}~\bibnamefont
  {Gilmour}},\ }\href {\doibase 10.1038/nrm2720} {\bibfield  {journal}
  {\bibinfo  {journal} {Nat Rev Mol Cell Biol}\ }\textbf {\bibinfo {volume}
  {10}},\ \bibinfo {pages} {445} (\bibinfo {year} {2009})}\BibitemShut
  {NoStop}%
\bibitem{Angelini2011}%
  \BibitemOpen
  \bibfield  {author} {\bibinfo {author} {\bibfnamefont {T.~E.}\ \bibnamefont
  {Angelini}}, \bibinfo {author} {\bibfnamefont {E.}~\bibnamefont {Hannezo}},
  \bibinfo {author} {\bibfnamefont {X.}~\bibnamefont {Trepat}}, \bibinfo
  {author} {\bibfnamefont {M.}~\bibnamefont {Marquez}}, \bibinfo {author}
  {\bibfnamefont {J.~J.}\ \bibnamefont {Fredberg}}, \ and\ \bibinfo {author}
  {\bibfnamefont {D.~A.}\ \bibnamefont {Weitz}},\ }\href {\doibase
  10.1073/pnas.1010059108} {\bibfield  {journal} {\bibinfo  {journal} {Proc.
  Natl. Acad. Sci. USA}\ }\textbf {\bibinfo {volume} {108}},\ \bibinfo {pages}
  {4714} (\bibinfo {year} {2011})}\BibitemShut {NoStop}%
\bibitem{Nnetu2012}%
  \BibitemOpen
  \bibfield  {author} {\bibinfo {author} {\bibfnamefont {K.~D.}\ \bibnamefont
  {Nnetu}}, \bibinfo {author} {\bibfnamefont {M.}~\bibnamefont {Knorr}},
  \bibinfo {author} {\bibfnamefont {J.}~\bibnamefont {K{\"{a}}s}}, \ and\
  \bibinfo {author} {\bibfnamefont {M.}~\bibnamefont {Zink}},\ }\href {\doibase 10.1088/1367-2630/14/11/115012} {\bibfield  {journal} {\bibinfo  {journal}
  {New J. Phys.}\ }\textbf {\bibinfo {volume} {14}},\ \bibinfo {pages} {115012}
  (\bibinfo {year} {2012})}\BibitemShut {NoStop}%
\bibitem{Park2015}%
  \BibitemOpen
  \bibfield  {author} {\bibinfo {author} {\bibfnamefont {J.~A.}\ \bibnamefont
  {Park}}, \bibinfo {author} {\bibfnamefont {J.~H.}\ \bibnamefont {Kim}},
  \bibinfo {author} {\bibfnamefont {D.}~\bibnamefont {Bi}}, \bibinfo {author}
  {\bibfnamefont {J.~A.}\ \bibnamefont {Mitchel}}, \bibinfo {author}
  {\bibfnamefont {N.~T.}\ \bibnamefont {Qazvini}}, \bibinfo {author}
  {\bibfnamefont {K.}~\bibnamefont {Tantisira}}, \bibinfo {author}
  {\bibfnamefont {C.~Y.}\ \bibnamefont {Park}}, \bibinfo {author}
  {\bibfnamefont {M.}~\bibnamefont {McGill}}, \bibinfo {author} {\bibfnamefont
  {S.~H.}\ \bibnamefont {Kim}}, \bibinfo {author} {\bibfnamefont
  {B.}~\bibnamefont {Gweon}}, \bibinfo {author} {\bibfnamefont
  {J.}~\bibnamefont {Notbohm}}, \bibinfo {author} {\bibfnamefont
  {R.}~\bibnamefont {Steward}}, \bibinfo {author} {\bibfnamefont
  {S.}~\bibnamefont {Burger}}, \bibinfo {author} {\bibfnamefont {S.~H.}\
  \bibnamefont {Randell}}, \bibinfo {author} {\bibfnamefont {A.~T.}\
  \bibnamefont {Kho}}, \bibinfo {author} {\bibfnamefont {D.~T.}\ \bibnamefont
  {Tambe}}, \bibinfo {author} {\bibfnamefont {C.}~\bibnamefont {Hardin}},
  \bibinfo {author} {\bibfnamefont {S.~A.}\ \bibnamefont {Shore}}, \bibinfo
  {author} {\bibfnamefont {E.}~\bibnamefont {Israel}}, \bibinfo {author}
  {\bibfnamefont {D.~A.}\ \bibnamefont {Weitz}}, \bibinfo {author}
  {\bibfnamefont {D.~J.}\ \bibnamefont {Tschumperlin}}, \bibinfo {author}
  {\bibfnamefont {E.~P.}\ \bibnamefont {Henske}}, \bibinfo {author}
  {\bibfnamefont {S.~T.}\ \bibnamefont {Weiss}}, \bibinfo {author}
  {\bibfnamefont {M.~L.}\ \bibnamefont {Manning}}, \bibinfo {author}
  {\bibfnamefont {J.~P.}\ \bibnamefont {Butler}}, \bibinfo {author}
  {\bibfnamefont {J.~M.}\ \bibnamefont {Drazen}}, \ and\ \bibinfo {author}
  {\bibfnamefont {J.~J.}\ \bibnamefont {Fredberg}},\ }\href {\doibase
  10.1038/nmat4357} {\bibfield  {journal} {\bibinfo  {journal} {Nat. Mater.}\
  }\textbf {\bibinfo {volume} {14}},\ \bibinfo {pages} {1040} (\bibinfo {year}
  {2015})}\BibitemShut {NoStop}%
\bibitem{Garcia2015}%
  \BibitemOpen
  \bibfield  {author} {\bibinfo {author} {\bibfnamefont {S.}~\bibnamefont
  {Garcia}}, \bibinfo {author} {\bibfnamefont {E.}~\bibnamefont {Hannezo}},
  \bibinfo {author} {\bibfnamefont {J.}~\bibnamefont {Elgeti}}, \bibinfo
  {author} {\bibfnamefont {J.~F.}\ \bibnamefont {Joanny}}, \bibinfo {author}
  {\bibfnamefont {P.}~\bibnamefont {Silberzan}}, \ and\ \bibinfo {author}
  {\bibfnamefont {N.~S.}\ \bibnamefont {Gov}},\ }\href {\doibase
  10.1073/pnas.1510973112} {\bibfield  {journal} {\bibinfo  {journal} {Proc.
  Natl. Acad. Sci. USA}\ }\textbf {\bibinfo {volume} {112}},\ \bibinfo {pages}
  {15314} (\bibinfo {year} {2015})}\BibitemShut {NoStop}%
\bibitem{Malinverno2017}%
  \BibitemOpen
  \bibfield  {author} {\bibinfo {author} {\bibfnamefont {C.}~\bibnamefont
  {Malinverno}}, \bibinfo {author} {\bibfnamefont {S.}~\bibnamefont
  {Corallino}}, \bibinfo {author} {\bibfnamefont {F.}~\bibnamefont {Giavazzi}},
  \bibinfo {author} {\bibfnamefont {M.}~\bibnamefont {Bergert}}, \bibinfo
  {author} {\bibfnamefont {Q.}~\bibnamefont {Li}}, \bibinfo {author}
  {\bibfnamefont {M.}~\bibnamefont {Leoni}}, \bibinfo {author} {\bibfnamefont
  {A.}~\bibnamefont {Disanza}}, \bibinfo {author} {\bibfnamefont
  {E.}~\bibnamefont {Frittoli}}, \bibinfo {author} {\bibfnamefont
  {A.}~\bibnamefont {Oldani}}, \bibinfo {author} {\bibfnamefont
  {E.}~\bibnamefont {Martini}}, \bibinfo {author} {\bibfnamefont
  {T.}~\bibnamefont {Lendenmann}}, \bibinfo {author} {\bibfnamefont
  {G.}~\bibnamefont {Deflorian}}, \bibinfo {author} {\bibfnamefont {G.~V.}\
  \bibnamefont {Beznoussenko}}, \bibinfo {author} {\bibfnamefont
  {D.}~\bibnamefont {Poulikakos}}, \bibinfo {author} {\bibfnamefont {K.~H.}\
  \bibnamefont {Ong}}, \bibinfo {author} {\bibfnamefont {M.}~\bibnamefont
  {Uroz}}, \bibinfo {author} {\bibfnamefont {X.}~\bibnamefont {Trepat}},
  \bibinfo {author} {\bibfnamefont {D.}~\bibnamefont {Parazzoli}}, \bibinfo
  {author} {\bibfnamefont {P.}~\bibnamefont {Maiuri}}, \bibinfo {author}
  {\bibfnamefont {W.}~\bibnamefont {Yu}}, \bibinfo {author} {\bibfnamefont
  {A.}~\bibnamefont {Ferrari}}, \bibinfo {author} {\bibfnamefont
  {R.}~\bibnamefont {Cerbino}}, \ and\ \bibinfo {author} {\bibfnamefont
  {G.}~\bibnamefont {Scita}},\ }\href {\doibase 10.1038/nmat4848} {\bibfield
  {journal} {\bibinfo  {journal} {Nat. Mater.}\ }\textbf {\bibinfo {volume}
  {16}},\ \bibinfo {pages} {587} (\bibinfo {year} {2017})}\BibitemShut
  {NoStop}%
\bibitem{Atia2018}%
  \BibitemOpen
  \bibfield  {author} {\bibinfo {author} {\bibfnamefont {L.}~\bibnamefont
  {Atia}}, \bibinfo {author} {\bibfnamefont {D.}~\bibnamefont {Bi}}, \bibinfo
  {author} {\bibfnamefont {Y.}~\bibnamefont {Sharma}}, \bibinfo {author}
  {\bibfnamefont {J.~A.}\ \bibnamefont {Mitchel}}, \bibinfo {author}
  {\bibfnamefont {B.}~\bibnamefont {Gweon}}, \bibinfo {author} {\bibfnamefont
  {S.}~\bibnamefont {A.~Koehler}}, \bibinfo {author} {\bibfnamefont {S.~J.}\
  \bibnamefont {DeCamp}}, \bibinfo {author} {\bibfnamefont {B.}~\bibnamefont
  {Lan}}, \bibinfo {author} {\bibfnamefont {J.~H.}\ \bibnamefont {Kim}},
  \bibinfo {author} {\bibfnamefont {R.}~\bibnamefont {Hirsch}}, \bibinfo
  {author} {\bibfnamefont {A.~F.}\ \bibnamefont {Pegoraro}}, \bibinfo {author}
  {\bibfnamefont {K.~H.}\ \bibnamefont {Lee}}, \bibinfo {author} {\bibfnamefont
  {J.~R.}\ \bibnamefont {Starr}}, \bibinfo {author} {\bibfnamefont {D.~A.}\
  \bibnamefont {Weitz}}, \bibinfo {author} {\bibfnamefont {A.~C.}\ \bibnamefont
  {Martin}}, \bibinfo {author} {\bibfnamefont {J.-A.}\ \bibnamefont {Park}},
  \bibinfo {author} {\bibfnamefont {J.~P.}\ \bibnamefont {Butler}}, \ and\
  \bibinfo {author} {\bibfnamefont {J.~J.}\ \bibnamefont {Fredberg}},\ }\href
  {\doibase 10.1038/s41567-018-0089-9} {\bibfield  {journal} {\bibinfo
  {journal} {Nat. Phys.}\ }\textbf {\bibinfo {volume} {14}},\ \bibinfo {pages}
  {613} (\bibinfo {year} {2018})}\BibitemShut {NoStop}%
\bibitem{Mongera2018}%
  \BibitemOpen
  \bibfield  {author} {\bibinfo {author} {\bibfnamefont {A.}~\bibnamefont
  {Mongera}}, \bibinfo {author} {\bibfnamefont {P.}~\bibnamefont {Rowghanian}},
  \bibinfo {author} {\bibfnamefont {H.~J.}\ \bibnamefont {Gustafson}}, \bibinfo
  {author} {\bibfnamefont {E.}~\bibnamefont {Shelton}}, \bibinfo {author}
  {\bibfnamefont {D.~A.}\ \bibnamefont {Kealhofer}}, \bibinfo {author}
  {\bibfnamefont {E.~K.}\ \bibnamefont {Carn}}, \bibinfo {author}
  {\bibfnamefont {F.}~\bibnamefont {Serwane}}, \bibinfo {author} {\bibfnamefont
  {A.~A.}\ \bibnamefont {Lucio}}, \bibinfo {author} {\bibfnamefont
  {J.}~\bibnamefont {Giammona}}, \ and\ \bibinfo {author} {\bibfnamefont
  {O.}~\bibnamefont {Camp{\`{a}}s}},\ }\href {\doibase
  10.1038/s41586-018-0479-2} {\bibfield  {journal} {\bibinfo  {journal}
  {Nature}\ }\textbf {\bibinfo {volume} {561}},\ \bibinfo {pages} {401}
  (\bibinfo {year} {2018})}\BibitemShut {NoStop}%
\bibitem{Graner1992}%
  \BibitemOpen
  \bibfield  {author} {\bibinfo {author} {\bibfnamefont {F.}~\bibnamefont
  {Graner}}\ and\ \bibinfo {author} {\bibfnamefont {J.~A.}\ \bibnamefont
  {Glazier}},\ }\href {\doibase 10.1103/PhysRevLett.69.2013} {\bibfield
  {journal} {\bibinfo  {journal} {Phys. Rev. Lett.}\ }\textbf {\bibinfo
  {volume} {69}},\ \bibinfo {pages} {2013} (\bibinfo {year}
  {1992})}\BibitemShut {NoStop}%
\bibitem{Nagai2001}%
  \BibitemOpen
  \bibfield  {author} {\bibinfo {author} {\bibfnamefont {T.}~\bibnamefont
  {Nagai}}\ and\ \bibinfo {author} {\bibfnamefont {H.}~\bibnamefont {Honda}},\
  }\href {\doibase 10.1080/13642810108205772} {\bibfield  {journal} {\bibinfo
  {journal} {Philos. Mag. B}\ }\textbf {\bibinfo {volume} {81}},\ \bibinfo
  {pages} {699} (\bibinfo {year} {2001})}\BibitemShut {NoStop}%
\bibitem{Staple2010}%
  \BibitemOpen
  \bibfield  {author} {\bibinfo {author} {\bibfnamefont {D.~B.}\ \bibnamefont
  {Staple}}, \bibinfo {author} {\bibfnamefont {R.}~\bibnamefont {Farhadifar}},
  \bibinfo {author} {\bibfnamefont {J.~C.}\ \bibnamefont {R{\"{o}}per}},
  \bibinfo {author} {\bibfnamefont {B.}~\bibnamefont {Aigouy}}, \bibinfo
  {author} {\bibfnamefont {S.}~\bibnamefont {Eaton}}, \ and\ \bibinfo {author}
  {\bibfnamefont {F.}~\bibnamefont {J{\"{u}}licher}},\ }\href {\doibase
  10.1140/epje/i2010-10677-0} {\bibfield  {journal} {\bibinfo  {journal} {Eur.
  Phys. J. E}\ }\textbf {\bibinfo {volume} {33}},\ \bibinfo {pages} {117}
  (\bibinfo {year} {2010})}\BibitemShut {NoStop}%
\bibitem{Fletcher2014}%
  \BibitemOpen
  \bibfield  {author} {\bibinfo {author} {\bibfnamefont {A.~G.}\ \bibnamefont
  {Fletcher}}, \bibinfo {author} {\bibfnamefont {M.}~\bibnamefont
  {Osterfield}}, \bibinfo {author} {\bibfnamefont {R.~E.}\ \bibnamefont
  {Baker}}, \ and\ \bibinfo {author} {\bibfnamefont {S.~Y.}\ \bibnamefont
  {Shvartsman}},\ }\href {\doibase 10.1016/j.bpj.2013.11.4498} {\bibfield
  {journal} {\bibinfo  {journal} {Biophys. J.}\ }\textbf {\bibinfo {volume}
  {106}},\ \bibinfo {pages} {2291} (\bibinfo {year} {2014})}\BibitemShut
  {NoStop}%
\bibitem{Bi2015}%
  \BibitemOpen
  \bibfield  {author} {\bibinfo {author} {\bibfnamefont {D.}~\bibnamefont
  {Bi}}, \bibinfo {author} {\bibfnamefont {J.~H.}\ \bibnamefont {Lopez}},
  \bibinfo {author} {\bibfnamefont {J.~M.}\ \bibnamefont {Schwarz}}, \ and\
  \bibinfo {author} {\bibfnamefont {M.~L.}\ \bibnamefont {Manning}},\ }\href
  {\doibase 10.1038/nphys3471} {\bibfield  {journal} {\bibinfo  {journal}
  {Nature Phyics}\ }\textbf {\bibinfo {volume} {11}},\ \bibinfo {pages} {1074}
  (\bibinfo {year} {2015})}\BibitemShut {NoStop}%
\bibitem{Li2014b}%
  \BibitemOpen
  \bibfield  {author} {\bibinfo {author} {\bibfnamefont {B.}~\bibnamefont
  {Li}}\ and\ \bibinfo {author} {\bibfnamefont {S.~X.}\ \bibnamefont {Sun}},\
  }\href {\doibase 10.1016/j.bpj.2014.08.006} {\bibfield  {journal} {\bibinfo
  {journal} {Biophys. J.}\ }\textbf {\bibinfo {volume} {107}},\ \bibinfo
  {pages} {1532} (\bibinfo {year} {2014})}\BibitemShut {NoStop}%
\bibitem{Bi2016}%
  \BibitemOpen
  \bibfield  {author} {\bibinfo {author} {\bibfnamefont {D.}~\bibnamefont
  {Bi}}, \bibinfo {author} {\bibfnamefont {X.}~\bibnamefont {Yang}}, \bibinfo
  {author} {\bibfnamefont {M.~C.}\ \bibnamefont {Marchetti}}, \ and\ \bibinfo
  {author} {\bibfnamefont {M.~L.}\ \bibnamefont {Manning}},\ }\href {\doibase
  10.1103/PhysRevX.6.021011} {\bibfield  {journal} {\bibinfo  {journal} {Phys.
  Rev. X}\ }\textbf {\bibinfo {volume} {6}},\ \bibinfo {pages} {021011}
  (\bibinfo {year} {2016})}\BibitemShut {NoStop}%
\bibitem{Teomy2018c}%
  \BibitemOpen
  \bibfield  {author} {\bibinfo {author} {\bibfnamefont {E.}~\bibnamefont
  {Teomy}}, \bibinfo {author} {\bibfnamefont {D.~A.}\ \bibnamefont {Kessler}},
  \ and\ \bibinfo {author} {\bibfnamefont {H.}~\bibnamefont {Levine}},\ }\href
  {\doibase 10.1103/PhysRevE.98.042418} {\bibfield  {journal} {\bibinfo
  {journal} {Phys. Rev. E}\ }\textbf {\bibinfo {volume} {98}},\ \bibinfo
  {pages} {042418} (\bibinfo {year} {2018})}\BibitemShut {NoStop}%
\bibitem{Yan2019}%
  \BibitemOpen
  \bibfield  {author} {\bibinfo {author} {\bibfnamefont {L.}~\bibnamefont
  {Yan}}\ and\ \bibinfo {author} {\bibfnamefont {D.}~\bibnamefont {Bi}},\
  }\href {\doibase 10.1103/PhysRevX.9.011029} {\bibfield  {journal} {\bibinfo
  {journal} {Phys. Rev. X}\ }\textbf {\bibinfo {volume} {9}},\ \bibinfo {pages}
  {011029} (\bibinfo {year} {2019})}\BibitemShut {NoStop}%
\bibitem{Li2018}%
  \BibitemOpen
  \bibfield  {author} {\bibinfo {author} {\bibfnamefont {Y.-W.}\ \bibnamefont
  {Li}}\ and\ \bibinfo {author} {\bibfnamefont {M.~P.}\ \bibnamefont
  {Ciamarra}},\ }\href {\doibase 10.1103/PhysRevMaterials.2.045602} {\bibfield
  {journal} {\bibinfo  {journal} {Phys. Rev. Mater.}\ }\textbf {\bibinfo
  {volume} {2}},\ \bibinfo {pages} {045602} (\bibinfo {year}
  {2018})}\BibitemShut {NoStop}%
\bibitem{Durand2019}%
  \BibitemOpen
  \bibfield  {author} {\bibinfo {author} {\bibfnamefont {M.}~\bibnamefont
  {Durand}}\ and\ \bibinfo {author} {\bibfnamefont {J.}~\bibnamefont {Heu}},\
  }\href {\doibase 10.1103/PhysRevLett.123.188001} {\bibfield  {journal}
  {\bibinfo  {journal} {Phys. Rev. Lett.}\ }\textbf {\bibinfo {volume} {123}},\
  \bibinfo {pages} {188001} (\bibinfo {year} {2019})}\BibitemShut {NoStop}%
\bibitem{Giavazzi2018}%
  \BibitemOpen
  \bibfield  {author} {\bibinfo {author} {\bibfnamefont {F.}~\bibnamefont
  {Giavazzi}}, \bibinfo {author} {\bibfnamefont {M.}~\bibnamefont {Paoluzzi}},
  \bibinfo {author} {\bibfnamefont {M.}~\bibnamefont {Macchi}}, \bibinfo
  {author} {\bibfnamefont {D.}~\bibnamefont {Bi}}, \bibinfo {author}
  {\bibfnamefont {G.}~\bibnamefont {Scita}}, \bibinfo {author} {\bibfnamefont
  {M.~L.}\ \bibnamefont {Manning}}, \bibinfo {author} {\bibfnamefont
  {R.}~\bibnamefont {Cerbino}}, \ and\ \bibinfo {author} {\bibfnamefont
  {M.~C.}\ \bibnamefont {Marchetti}},\ }\href {\doibase 10.1039/c8sm00126j}
  {\bibfield  {journal} {\bibinfo  {journal} {Soft Matter}\ }\textbf {\bibinfo
  {volume} {14}},\ \bibinfo {pages} {3471} (\bibinfo {year}
  {2018})}\BibitemShut {NoStop}%
\bibitem{Barton2017}%
  \BibitemOpen
  \bibfield  {author} {\bibinfo {author} {\bibfnamefont {D.~L.}\ \bibnamefont
  {Barton}}, \bibinfo {author} {\bibfnamefont {S.}~\bibnamefont {Henkes}},
  \bibinfo {author} {\bibfnamefont {C.~J.}\ \bibnamefont {Weijer}}, \ and\
  \bibinfo {author} {\bibfnamefont {R.}~\bibnamefont {Sknepnek}},\ }\href
  {\doibase 10.1371/journal.pcbi.1005569} {\bibfield  {journal} {\bibinfo
  {journal} {PLoS Comput. Biol.}\ }\textbf {\bibinfo {volume} {13}},\
  (\bibinfo {year} {2017})}\BibitemShut {NoStop}%
\bibitem{Chiang2016}%
  \BibitemOpen
  \bibfield  {author} {\bibinfo {author} {\bibfnamefont {M.}~\bibnamefont
  {Chiang}}\ and\ \bibinfo {author} {\bibfnamefont {D.}~\bibnamefont
  {Marenduzzo}},\ }\href {\doibase 10.1209/0295-5075/116/28009} {\bibfield
  {journal} {\bibinfo  {journal} {EPL}\ }\textbf {\bibinfo {volume} {116}},\
  \bibinfo {pages} {28009} (\bibinfo {year} {2016})}\BibitemShut {NoStop}%
\bibitem{Engel2013}%
  \BibitemOpen
  \bibfield  {author} {\bibinfo {author} {\bibfnamefont {M.}~\bibnamefont
  {Engel}}, \bibinfo {author} {\bibfnamefont {J.~A.}\ \bibnamefont {Anderson}},
  \bibinfo {author} {\bibfnamefont {S.~C.}\ \bibnamefont {Glotzer}}, \bibinfo
  {author} {\bibfnamefont {M.}~\bibnamefont {Isobe}}, \bibinfo {author}
  {\bibfnamefont {E.~P.}\ \bibnamefont {Bernard}}, \ and\ \bibinfo {author}
  {\bibfnamefont {W.}~\bibnamefont {Krauth}},\ }\href {\doibase
  10.1103/PhysRevE.87.042134} {\bibfield  {journal} {\bibinfo  {journal} {Phys.
  Rev. E}\ }\textbf {\bibinfo {volume} {87}},\ \bibinfo {pages} {042134}
  (\bibinfo {year} {2013})}\BibitemShut {NoStop}%
\bibitem{Kapfer2015}%
  \BibitemOpen
  \bibfield  {author} {\bibinfo {author} {\bibfnamefont {S.~C.}\ \bibnamefont
  {Kapfer}}\ and\ \bibinfo {author} {\bibfnamefont {W.}~\bibnamefont
  {Krauth}},\ }\href {\doibase 10.1103/PhysRevLett.114.035702} {\bibfield
  {journal} {\bibinfo  {journal} {Phys. Rev. Lett.}\ }\textbf {\bibinfo
  {volume} {114}},\ \bibinfo {pages} {035702} (\bibinfo {year}
  {2015})}\BibitemShut {NoStop}%
\bibitem{Hajibabaei2019}%
  \BibitemOpen
  \bibfield  {author} {\bibinfo {author} {\bibfnamefont {A.}~\bibnamefont
  {Hajibabaei}}\ and\ \bibinfo {author} {\bibfnamefont {K.~S.}\ \bibnamefont
  {Kim}},\ }\href {\doibase 10.1103/PhysRevE.99.022145} {\bibfield  {journal}
  {\bibinfo  {journal} {Phys. Rev. E}\ }\textbf {\bibinfo {volume} {99}},\
  \bibinfo {pages} {022145} (\bibinfo {year} {2019})}\BibitemShut {NoStop}%
\bibitem{Digregorio2019}%
  \BibitemOpen
  \bibfield  {author} {\bibinfo {author} {\bibfnamefont {P.}~\bibnamefont
  {Digregorio}}, \bibinfo {author} {\bibfnamefont {D.}~\bibnamefont {Levis}},
  \bibinfo {author} {\bibfnamefont {L.~F.}\ \bibnamefont {Cugliandolo}},
  \bibinfo {author} {\bibfnamefont {G.}~\bibnamefont {Gonnella}}, \ and\
  \bibinfo {author} {\bibfnamefont {I.}~\bibnamefont {Pagonabarraga}},\ }\href
  {https://arxiv.org/abs/1911.06366} {\bibfield  {journal} {\bibinfo  {journal}
  {arXiv:1911.06366}\ } (\bibinfo {year} {2019})}\BibitemShut {NoStop}%
\bibitem{Saito1982}%
  \BibitemOpen
  \bibfield  {author} {\bibinfo {author} {\bibfnamefont {Y.}~\bibnamefont
  {Saito}},\ }\href {\doibase 10.1103/PhysRevLett.48.1114} {\bibfield
  {journal} {\bibinfo  {journal} {Phys. Rev. Lett.}\ }\textbf {\bibinfo
  {volume} {48}},\ \bibinfo {pages} {1114} (\bibinfo {year}
  {1982})}\BibitemShut {NoStop}%
\bibitem{Chui1983}%
  \BibitemOpen
  \bibfield  {author} {\bibinfo {author} {\bibfnamefont {S.~T.}\ \bibnamefont
  {Chui}},\ }\href {\doibase 10.1103/PhysRevB.28.178} {\bibfield  {journal}
  {\bibinfo  {journal} {Phys. Rev. B}\ }\textbf {\bibinfo {volume} {28}},\
  \bibinfo {pages} {178} (\bibinfo {year} {1983})}\BibitemShut {NoStop}%
\bibitem{Kosterlitz1973}%
  \BibitemOpen
  \bibfield  {author} {\bibinfo {author} {\bibfnamefont {J.~M.}\ \bibnamefont
  {Kosterlitz}}\ and\ \bibinfo {author} {\bibfnamefont {D.~J.}\ \bibnamefont
  {Thouless}},\ }\href {\doibase 10.1088/0022-3719/6/7/010} {\bibfield
  {journal} {\bibinfo  {journal} {J. Phys. C Solid State Phys.}\ }\textbf
  {\bibinfo {volume} {6}},\ \bibinfo {pages} {1181} (\bibinfo {year}
  {1973})}\BibitemShut {NoStop}%
\bibitem{Halperin1978}%
  \BibitemOpen
  \bibfield  {author} {\bibinfo {author} {\bibfnamefont {B.~I.}\ \bibnamefont
  {Halperin}}\ and\ \bibinfo {author} {\bibfnamefont {D.~R.}\ \bibnamefont
  {Nelson}},\ }\href {\doibase 10.1103/PhysRevLett.41.121} {\bibfield
  {journal} {\bibinfo  {journal} {Phys. Rev. Lett.}\ }\textbf {\bibinfo
  {volume} {41}},\ \bibinfo {pages} {121} (\bibinfo {year} {1978})}\BibitemShut
  {NoStop}%
\bibitem{Nelson1978}%
  \BibitemOpen
  \bibfield  {author} {\bibinfo {author} {\bibfnamefont {D.~R.}\ \bibnamefont
  {Nelson}},\ }\href {\doibase 10.1103/PhysRevB.18.2318} {\bibfield  {journal}
  {\bibinfo  {journal} {Phys. Rev. B}\ }\textbf {\bibinfo {volume} {18}},\
  \bibinfo {pages} {2318} (\bibinfo {year} {1978})}\BibitemShut {NoStop}%
\bibitem{Young1979}%
  \BibitemOpen
  \bibfield  {author} {\bibinfo {author} {\bibfnamefont {A.~P.}\ \bibnamefont
  {Young}},\ }\href {\doibase 10.1103/PhysRevB.19.1855} {\bibfield  {journal}
  {\bibinfo  {journal} {Phys. Rev. B}\ }\textbf {\bibinfo {volume} {19}},\
  \bibinfo {pages} {1855} (\bibinfo {year} {1979})}\BibitemShut {NoStop}%
\bibitem{Dash1999}%
  \BibitemOpen
  \bibfield  {author} {\bibinfo {author} {\bibfnamefont {J.~G.}\ \bibnamefont
  {Dash}},\ }\href {\doibase 10.1103/revmodphys.71.1737} {\bibfield  {journal}
  {\bibinfo  {journal} {Rev. Mod. Phys.}\ }\textbf {\bibinfo {volume} {71}},\
  \bibinfo {pages} {1737} (\bibinfo {year} {1999})}\BibitemShut {NoStop}%
\bibitem{Gasser2009}%
  \BibitemOpen
  \bibfield  {author} {\bibinfo {author} {\bibfnamefont {U.}~\bibnamefont
  {Gasser}},\ }\href {\doibase 10.1088/0953-8984/21/20/203101} {\bibfield
  {journal} {\bibinfo  {journal} {J. Phys. Condens. Matter}\ }\textbf {\bibinfo
  {volume} {21}},\  (\bibinfo {year} {2009})}\BibitemShut {NoStop}%
\bibitem{Etournay2015}%
  \BibitemOpen
  \bibfield  {author} {\bibinfo {author} {\bibfnamefont {R.}~\bibnamefont
  {Etournay}}, \bibinfo {author} {\bibfnamefont {M.}~\bibnamefont
  {Popovi{\'{c}}}}, \bibinfo {author} {\bibfnamefont {M.}~\bibnamefont
  {Merkel}}, \bibinfo {author} {\bibfnamefont {A.}~\bibnamefont {Nandi}},
  \bibinfo {author} {\bibfnamefont {C.}~\bibnamefont {Blasse}}, \bibinfo
  {author} {\bibfnamefont {B.}~\bibnamefont {Aigouy}}, \bibinfo {author}
  {\bibfnamefont {H.}~\bibnamefont {Brandl}}, \bibinfo {author} {\bibfnamefont
  {G.}~\bibnamefont {Myers}}, \bibinfo {author} {\bibfnamefont
  {G.}~\bibnamefont {Salbreux}}, \bibinfo {author} {\bibfnamefont
  {F.}~\bibnamefont {J{\"{u}}licher}}, \ and\ \bibinfo {author} {\bibfnamefont
  {S.}~\bibnamefont {Eaton}},\ }\href {\doibase 10.7554/eLife.07090} {\bibfield
   {journal} {\bibinfo  {journal} {Elife}\ }\textbf {\bibinfo {volume} {4}},\
  \bibinfo {pages} {e07090} (\bibinfo {year} {2015})}\BibitemShut {NoStop}%
\bibitem{Eisenhoffer2012}%
  \BibitemOpen
  \bibfield  {author} {\bibinfo {author} {\bibfnamefont {G.~T.}\ \bibnamefont
  {Eisenhoffer}}, \bibinfo {author} {\bibfnamefont {P.~D.}\ \bibnamefont
  {Loftus}}, \bibinfo {author} {\bibfnamefont {M.}~\bibnamefont {Yoshigi}},
  \bibinfo {author} {\bibfnamefont {H.}~\bibnamefont {Otsuna}}, \bibinfo
  {author} {\bibfnamefont {C.~B.}\ \bibnamefont {Chien}}, \bibinfo {author}
  {\bibfnamefont {P.~A.}\ \bibnamefont {Morcos}}, \ and\ \bibinfo {author}
  {\bibfnamefont {J.}~\bibnamefont {Rosenblatt}},\ }\href {\doibase
  10.1038/nature10999} {\bibfield  {journal} {\bibinfo  {journal} {Nature}\
  }\textbf {\bibinfo {volume} {484}},\ \bibinfo {pages} {546} (\bibinfo {year}
  {2012})}\BibitemShut {NoStop}%
\bibitem{Koster2007}%
  \BibitemOpen
  \bibfield  {author} {\bibinfo {author} {\bibfnamefont {M.~I.}\ \bibnamefont
  {Koster}}\ and\ \bibinfo {author} {\bibfnamefont {D.~R.}\ \bibnamefont
  {Roop}},\ }\href {\doibase 10.1146/annurev.cellbio.23.090506.123357}
  {\bibfield  {journal} {\bibinfo  {journal} {Annu. Rev. Cell Dev. Biol.}\
  }\textbf {\bibinfo {volume} {23}},\ \bibinfo {pages} {93} (\bibinfo {year}
  {2007})}\BibitemShut {NoStop}%
\bibitem{Nonomura2012}%
  \BibitemOpen
  \bibfield  {author} {\bibinfo {author} {\bibfnamefont {M.}~\bibnamefont
  {Nonomura}},\ }\href {\doibase 10.1371/journal.pone.0033501} {\bibfield
  {journal} {\bibinfo  {journal} {PLoS One}\ }\textbf {\bibinfo {volume} {7}},\
  \bibinfo {pages} {e33501} (\bibinfo {year} {2012})}\BibitemShut {NoStop}%
\bibitem{Palmieri2015}%
  \BibitemOpen
  \bibfield  {author} {\bibinfo {author} {\bibfnamefont {B.}~\bibnamefont
  {Palmieri}}, \bibinfo {author} {\bibfnamefont {Y.}~\bibnamefont {Bresler}},
  \bibinfo {author} {\bibfnamefont {D.}~\bibnamefont {Wirtz}}, \ and\ \bibinfo
  {author} {\bibfnamefont {M.}~\bibnamefont {Grant}},\ }\href {\doibase
  10.1038/srep11745} {\bibfield  {journal} {\bibinfo  {journal} {Sci. Rep.}\
  }\textbf {\bibinfo {volume} {5}},\ \bibinfo {pages} {11745} (\bibinfo {year}
  {2015})}\BibitemShut {NoStop}%
\bibitem{Foglino2017}%
  \BibitemOpen
  \bibfield  {author} {\bibinfo {author} {\bibfnamefont {M.}~\bibnamefont
  {Foglino}}, \bibinfo {author} {\bibfnamefont {A.~N.}\ \bibnamefont
  {Morozov}}, \bibinfo {author} {\bibfnamefont {O.}~\bibnamefont {Henrich}}, \
  and\ \bibinfo {author} {\bibfnamefont {D.}~\bibnamefont {Marenduzzo}},\
  }\href {\doibase 10.1103/PhysRevLett.119.208002} {\bibfield  {journal}
  {\bibinfo  {journal} {Phys. Rev. Lett.}\ }\textbf {\bibinfo {volume} {119}},\
  \bibinfo {pages} {208002} (\bibinfo {year} {2017})}\BibitemShut {NoStop}%
\bibitem{Mueller2019}%
  \BibitemOpen
  \bibfield  {author} {\bibinfo {author} {\bibfnamefont {R.}~\bibnamefont
  {Mueller}}, \bibinfo {author} {\bibfnamefont {J.~M.}\ \bibnamefont
  {Yeomans}}, \ and\ \bibinfo {author} {\bibfnamefont {A.}~\bibnamefont
  {Doostmohammadi}},\ }\href {\doibase 10.1103/PhysRevLett.122.048004}
  {\bibfield  {journal} {\bibinfo  {journal} {Phys. Rev. Lett.}\ }\textbf
  {\bibinfo {volume} {122}},\ \bibinfo {pages} {048004} (\bibinfo {year}
  {2019})}\BibitemShut {NoStop}%
\bibitem{Pagonabarraga2002}%
  \BibitemOpen
  \bibfield  {author} {\bibinfo {author} {\bibfnamefont {I.}~\bibnamefont
  {Pagonabarraga}}, \bibinfo {author} {\bibfnamefont {A.~J.}\ \bibnamefont
  {Wagner}}, \ and\ \bibinfo {author} {\bibfnamefont {M.~E.}\ \bibnamefont
  {Cates}},\ }\href {\doibase 10.1023/A:1014594101067} {\bibfield  {journal}
  {\bibinfo  {journal} {J. Stat. Phys.}\ }\textbf {\bibinfo {volume} {107}},\
  \bibinfo {pages} {39} (\bibinfo {year} {2002})}\BibitemShut {NoStop}%
\bibitem{SM}%
  \BibitemOpen
  \href@noop {} {}\bibinfo {note} {See online supplemental material, which
  contains additional details on the model and additional results complementing
  those shown in the main text.}\BibitemShut {Stop}%
\bibitem{Romanczuk2012}%
  \BibitemOpen
  \bibfield  {author} {\bibinfo {author} {\bibfnamefont {P.}~\bibnamefont
  {Romanczuk}}, \bibinfo {author} {\bibfnamefont {M.}~\bibnamefont {B{\"a}r}},
  \bibinfo {author} {\bibfnamefont {W.}~\bibnamefont {Ebeling}}, \bibinfo
  {author} {\bibfnamefont {B.}~\bibnamefont {Lindner}}, \ and\ \bibinfo
  {author} {\bibfnamefont {L.}~\bibnamefont {Schimansky-Geier}},\ }\href
  {\doibase 10.1140/epjst/e2012-01529-y} {\bibfield  {journal} {\bibinfo
  {journal} {Eur. Phys. J. Spec. Top.}\ }\textbf {\bibinfo {volume} {202}},\
  \bibinfo {pages} {1} (\bibinfo {year} {2012})}\BibitemShut {NoStop}%
\bibitem{Fily2012}%
  \BibitemOpen
  \bibfield  {author} {\bibinfo {author} {\bibfnamefont {Y.}~\bibnamefont
  {Fily}}\ and\ \bibinfo {author} {\bibfnamefont {M.~C.}\ \bibnamefont
  {Marchetti}},\ }\href {\doibase 10.1103/PhysRevLett.108.235702} {\bibfield
  {journal} {\bibinfo  {journal} {Phys. Rev. Lett.}\ }\textbf {\bibinfo
  {volume} {108}},\ \bibinfo {pages} {235702} (\bibinfo {year}
  {2012})}\BibitemShut {NoStop}%
\bibitem{Fily2014}%
  \BibitemOpen
  \bibfield  {author} {\bibinfo {author} {\bibfnamefont {Y.}~\bibnamefont
  {Fily}}, \bibinfo {author} {\bibfnamefont {S.}~\bibnamefont {Henkes}}, \ and\
  \bibinfo {author} {\bibfnamefont {M.~C.}\ \bibnamefont {Marchetti}},\ }\href
  {\doibase 10.1039/c3sm52469h} {\bibfield  {journal} {\bibinfo  {journal}
  {Soft Matter}\ }\textbf {\bibinfo {volume} {10}},\ \bibinfo {pages} {2132}
  (\bibinfo {year} {2014})}\BibitemShut {NoStop}%
\bibitem{Kawaguchi2017}%
  \BibitemOpen
  \bibfield  {author} {\bibinfo {author} {\bibfnamefont {K.}~\bibnamefont
  {Kawaguchi}}, \bibinfo {author} {\bibfnamefont {R.}~\bibnamefont {Kageyama}},
  \ and\ \bibinfo {author} {\bibfnamefont {M.}~\bibnamefont {Sano}},\ }\href
  {\doibase 10.1038/nature22321} {\bibfield  {journal} {\bibinfo  {journal}
  {Nature}\ }\textbf {\bibinfo {volume} {545}},\ \bibinfo {pages} {327}
  (\bibinfo {year} {2017})}\BibitemShut {NoStop}%
\bibitem{Saw2017}%
  \BibitemOpen
  \bibfield  {author} {\bibinfo {author} {\bibfnamefont {T.~B.}\ \bibnamefont
  {Saw}}, \bibinfo {author} {\bibfnamefont {A.}~\bibnamefont {Doostmohammadi}},
  \bibinfo {author} {\bibfnamefont {V.}~\bibnamefont {Nier}}, \bibinfo {author}
  {\bibfnamefont {L.}~\bibnamefont {Kocgozlu}}, \bibinfo {author}
  {\bibfnamefont {S.}~\bibnamefont {Thampi}}, \bibinfo {author} {\bibfnamefont
  {Y.}~\bibnamefont {Toyama}}, \bibinfo {author} {\bibfnamefont
  {P.}~\bibnamefont {Marcq}}, \bibinfo {author} {\bibfnamefont {C.~T.}\
  \bibnamefont {Lim}}, \bibinfo {author} {\bibfnamefont {J.~M.}\ \bibnamefont
  {Yeomans}}, \ and\ \bibinfo {author} {\bibfnamefont {B.}~\bibnamefont
  {Ladoux}},\ }\href {\doibase 10.1038/nature21718} {\bibfield  {journal}
  {\bibinfo  {journal} {Nature}\ }\textbf {\bibinfo {volume} {544}},\ \bibinfo
  {pages} {212} (\bibinfo {year} {2017})}\BibitemShut {NoStop}%
\bibitem{Bowick2009}%
  \BibitemOpen
  \bibfield  {author} {\bibinfo {author} {\bibfnamefont {M.~J.}\ \bibnamefont
  {Bowick}}\ and\ \bibinfo {author} {\bibfnamefont {L.}~\bibnamefont {Giomi}},\
  }\href {\doibase 10.1080/00018730903043166} {\bibfield  {journal} {\bibinfo
  {journal} {Adv. Phys.}\ }\textbf {\bibinfo {volume} {58}},\ \bibinfo {pages}
  {449} (\bibinfo {year} {2009})}\BibitemShut {NoStop}%
\bibitem{Digregorio2018}%
  \BibitemOpen
  \bibfield  {author} {\bibinfo {author} {\bibfnamefont {P.}~\bibnamefont
  {Digregorio}}, \bibinfo {author} {\bibfnamefont {D.}~\bibnamefont {Levis}},
  \bibinfo {author} {\bibfnamefont {A.}~\bibnamefont {Suma}}, \bibinfo {author}
  {\bibfnamefont {L.~F.}\ \bibnamefont {Cugliandolo}}, \bibinfo {author}
  {\bibfnamefont {G.}~\bibnamefont {Gonnella}}, \ and\ \bibinfo {author}
  {\bibfnamefont {I.}~\bibnamefont {Pagonabarraga}},\ }\href {\doibase
  10.1103/PhysRevLett.121.098003} {\bibfield  {journal} {\bibinfo  {journal}
  {Phys. Rev. Lett.}\ }\textbf {\bibinfo {volume} {121}},\ \bibinfo {pages}
  {098003} (\bibinfo {year} {2018})}\BibitemShut {NoStop}%
\end{thebibliography}
\end{document}


\title{Solid-Liquid Transition of Deformable and Overlapping Active Particles: \\ Supplemental Material}

\author{Benjamin Loewe}
\affiliation{\ucsb}
\author{Michael Chiang}
\affiliation{\ue}
\author{Davide Marenduzzo}
\affiliation{\ue}
\author{M. Cristina Marchetti}
\affiliation{\ucsb}

\maketitle

\section{PHASE FIELD DYNAMICS}
In our study we consider a dry system of deformable particles, described as interacting phase fields with the solvent only providing friction. The particles are self-propelled and interact with each other via a repulsive interaction. Our particles are essentially deformable active Brownian particles (ABP) capable of overlap.
The dynamics of the $i$th phase field is governed by the equation
\begin{equation}
\label{eq:EOM}
\pd_t\phi_i+\vec{v}_i^a\cdot\vec{\nabla}\phi_i=-\frac{1}{\gamma}\frac{\delta \mathcal{F}}{\delta\phi_i},
\end{equation}
where $\mathcal{F}$ is the cells' equilibrium free energy 
\begin{equation}
\begin{split}
\F = \sum_{i=1}^{N}\Bigg[&\int\fd^2\vec{r}\,\left(\frac{\alpha}{4}\phi_i^2\left(\phi_i-\phi_0\right)^2+\frac{K}{2}\left(\vec{\nabla}\phi_i\right)^2\right)\\ &+\lambda\delta V[\phi_i]^2+\varepsilon\sum_{i<j=1}^{N}\int\fd^2\vec{r}\,\phi_i^2\phi_j^2\Bigg], \label{eqn:free_energy}
\end{split}
\end{equation}
with
\begin{align}
\delta V[\phi_i] = 1-\int\fd^2\vec{r}\,\frac{\phi_i^2}{\pi R^2\phi_0^2}.
\end{align}
Here, we discuss the physical origin of the advective velocity $\vec{v}_i^a$, where an additional superscript $a$ has been introduced for clarity (but it is dropped in the main paper).
In particular, we would like to contrast $\vec{v}_i^a$ to the center of mass velocity of the cell, defined as
\begin{equation}
    \vec{v}_i^{\text{cm}} = \frac{\fd}{\fd t} \vec{R}^{\text{cm}}_i,
\end{equation}
where 
\begin{equation}\label{CM}
    \vec{R}^{\text{cm}}_i \equiv \frac{1}{M_i} \int\fd^2\vec{r}\,\vec{r}\phi_i
\end{equation}
and $M_i$ is the cell's total ``mass'', 
\begin{equation}
    M_i \equiv \int \fd^2\vec{r}\,\phi_i.
\end{equation}

\subsection{Single Cell}
For a single isolated cell in the absence of external driving forces, it is easy to show that $\vec{v}^\text{cm}_i=\vec{v}_i^a$, as all forces acting on the cell are internal and cancel out. This is seen explicitly by taking the time derivative of Eq.~(\ref{CM})
\begin{align}
\begin{split}
  \vec{v}_i^{\text{cm}} = \frac{1}{M_i}\int&\fd^2\vec{r}\,\vec{r}(\pd_t \phi_i)
  - \frac{\vec{R}_i^{\text{cm}}}{M_i}\int\fd^2\vec{r}\,(\pd_t\phi_i),
\end{split}
\end{align}
which, using Eq.~(\ref{eq:EOM}), can be rewritten as
\begin{align}
\begin{split}
   M_i \vec{v}_i^{\text{cm}}= &-\int\fd^2\vec{r}\,\vec{r}(\vec{v}^a_{i}\cdot\vec{\nabla}\phi_i) + \vec{R}^{\text{cm}}_i\int\fd^2\vec{r}\,\vec{v}^a_{i}\cdot\vec{\nabla}\phi_i \\
    &- \frac{1}{\gamma}\int\fd^2\vec{r}\,\left(\vec{r}-\vec{R}^{\text{cm}}_i\right)\frac{\delta\F}{\delta\phi_i}.
\end{split}
\end{align}
Integrating the first two terms by parts, replacing the explicit form of the free energy, Eq.~(\ref{eqn:free_energy}), but without interactions, and changing variables to coordinates relative to the particle's center of mass, $\vec{u}=\vec{r}-\vec{R}_i^{\text{cm}}$, we obtain
\begin{align}
\begin{split}
    \vec{v}_i^{\text{cm}} =\;& \vec{v}^a_i - \frac{1}{\gamma M_i}\int\fd^2\vec{u}\,\vec{u}\bigg[\alpha\bigg(\phi_i^3-\frac{3}{2}\phi_{0}\phi_i^2+\frac{1}{2}\phi^2_{0}\phi_i\bigg)\\
     & -K\nabla^2\phi_i-\frac{4\lambda\phi_i}{\pi R^2\phi_{0}^2}\delta V[\phi_i]\bigg].\label{vcm1}
\end{split}
\end{align}
Assuming that the field $\phi_i$ is approximately uniform in the interior of the cell, we can write
\begin{equation}
    \int\fd^2\vec{u}\,\vec{u}\phi_i^3 \approx \int d^2\vec{u}\,\vec{u}\phi_i^2 \approx \int\fd^2\vec{u}\,\vec{u}\phi_i = 0,
\end{equation}
thus allowing us to neglect the first term in the integral in Eq.~(\ref{vcm1}). The third term also vanishes by the same reasoning. Similarly, given that $\phi_i$ is localized, we have
\begin{equation}
    \int\fd^2\vec{u}\,\vec{u}\nabla^2\phi_i = 0,
\end{equation}
and the second term also vanishes. Hence, for a single isolated cell we find
\begin{equation}
\vec{v}^a_i=\vec{v}^\text{cm}_i.\label{eq:v-one}
\end{equation}
In other words, for an isolated cell the advective velocity  in Eq.~(\ref{eq:EOM})  coincides with the center of mass velocity of the cell. If no externally applied nor internally generated (such as motility) forces act on the cell, then $\vec{v}^a_i=\vec{v}^\text{cm}_i=0$. In contrast, if the cell is active and self-motile, then $\vec{v}^a_i=\vec{v}^\text{cm}_i=\vec{v}_i^{\text{SP}}$, where  $\vec{v}_i^{\text{SP}}=v_0\,\hat{\bm e}$ is the cell motility, modeled as a propulsive velocity of constant speed $v_0$ and direction $\hat{\bm e}_i$ randomly rotated by noise, as described in the main paper.

\subsection{Interacting Cells}
In the presence of interactions, the procedure described above yields
\begin{equation}
    \vec{v}_i^{\text{cm}} = \vec{v}_i^a - \frac{2\varepsilon}{\gamma M_i}\sum_{j\neq i} \int\fd^2\vec{u}\,\vec{u}\phi_i\phi_j^2.\label{vcm_final}
\end{equation}
The second term on the right hand side of Eq.~(\ref{vcm_final}) arises from the forces acting on cell $i$ due to all other cells. Eq.~(\ref{vcm_final}) is a statement of force balance and can be rewritten as
\begin{equation}
\gamma \vec{v}_i^{\text{cm}} = \vec{F}_i^{\text{a}} +\vec{F}_i^{\text{int}},\label{force-balance}   
\end{equation}
 where the friction coefficient is assumed equal to the one in Eq.~(\ref{eq:EOM}) and
\begin{align}
&\vec{F}_i^a=\gamma\vec{v}_i^a,\\ \label{force_int}
&\vec{F}_i^{\text{int}} = -\frac{2\varepsilon}{M_i}\sum_{j\neq i} \int\fd^2\vec{u}\,\vec{u}\phi_i\phi_j^2.
\end{align}
It is evident that the center of mass velocity is indeed determined by both self-propulsion $\vec{v}_i^a=\vec{v}_i^{\text{SP}}$ and interactions. Incorporating interactions in the advective velocity $\vec{v}_i^a$ as done in much of the literature would lead to double counting for our dry system. 

The situation changes, however, if we assume the tissue to be surrounded by a solvent at low Reynolds number. In this case, the cell is not only advected by its self-propulsion but also by the velocity field of the solvent. Assuming that the solvent is in mechanical equilibrium with the cells, then the solvent velocity field must be equal to the gradient of the cell's stress tensor.  In this case, fluid-mediated interactions do lead to advection and $\vec{v}_i^a$ must include this contribution, as discussed in the literature~\cite{Palmieri2015,Mueller2019}.

\section{MODEL IMPLEMENTATIONS}
Here we describe the details of our numerical solution of Eq.~(\ref{eq:EOM}). The  full list of parameter values used in the simulations is provided in Table~S\ref{table:parameter_list}.
Evaluating the functional derivative of $\F$, Eq.~(\ref{eq:EOM}) can be written as
\begin{equation}
\begin{split}
  \pd_t\phi_i =& -\vec{v}_i^a\cdot\vec{\nabla}\phi_i -\frac{1}{\gamma}\bigg[\alpha\left(\phi_i^3-\frac{3}{2}\phi_i^2\phi_0+\frac{1}{2}\phi_i\phi_0^2\right)\\
    & - K\nabla^2\phi_i - \frac{4\lambda\phi_i}{\pi R^2\phi_0^2}\delta V[\phi_i] + 2\varepsilon\phi_i\left(h(\vec{r})-\phi_i^2\right)\bigg],\label{eqn:eom}
\end{split}
\end{equation}
where we have introduced an auxiliary field $h({\vec{r}}) = \sum_{i=1}^{N}\phi_i^2(\vec{r})$. This auxiliary field enables one to decouple and parallelize the computation of individual phase fields~\cite{Nonomura2012}. More precisely, we first calculate $h(\vec{r})$ using the phase fields at the current timestep, and then perform the update of individual phase fields in parallel with the knowledge of $h(\vec{r})$. 

We simulate Eq.~(\ref{eqn:eom}) using a finite difference method. Length is expressed in terms of the lattice spacing $\delta x$, and time in simulation time unit $\delta t$. For numerical stability, we set each timestep $\Delta t$ to $0.5\,\delta t$. The simulation code is written in a mixture of \texttt{C} and \texttt{C++} and is parallelized using \texttt{OpenMP}. In line with previous work~\cite{Nonomura2012,Mueller2019}, we use domain decomposition to improve computational efficiency. Each field is discretized as a square lattice with linear dimension $\ell_s = 41$ (which is much larger than the target radius of each cell, $R = 12$, but much smaller than the whole lattice), and we solve Eq.~(\ref{eqn:eom}) using fixed boundary conditions (i.e., $\phi_i = 0$) in these subdomains. Note that the boundary conditions for the full lattice are still periodic. During the simulations, we keep each cell near the center of its subdomain by performing a shifting algorithm which moves the cell back to the center of the subdomain when it has moved more than two lattice units in any direction. At the same time, we store reference coordinates of each cell relative to the full lattice, and these are updated accordingly when a shift has been performed.

\begin{table}
\begin{tabular}{|>{\centering}p{2cm}|>{\centering}p{2cm}|p{2cm}<{\centering}|}\hline
 Parameter & Dimensions & Value(s) \\\hline
 $\alpha$ & $E/L^2$ & $0.025$--$1.0$\\
 $K$ & $E$ & $0.05$--$2.0$\\
 $\lambda$ & $E$ & $6000$\\
$\varepsilon$ & $E/L^2$ & $0.1$\\
$\phi_0$ & $\text{-}$ & $2$\\
$\xi$ & $L$ & $2$\\
$R$ & $L$ & $12$\\
$\gamma$ & $ET/L^2$ & $10$\\
$D_r$ & $1/T$ & $0.0001$ \\
$v_0$ & $L/T$ & $0$--$0.0036$\\\hline
\end{tabular}
\caption{Parameter dimensions and values used in the simulation model. These values are in simulation units of length $\delta x$, time $\delta t$, and energy $\delta E$. Notice that the actual scale of energy is irrelevant as it scales out of Eq.~(\ref{eq:EOM}).}
\label{table:parameter_list}
\end{table}

We consider both a system of $N = 36$ and $100$ cells for most of our analysis~\footnote{Unless otherwise stated, the figures shown in this document are for the system size $N = 100$.}. We initialize the cells in a hexagonal arrangement, which is achieved by placing a circular droplet of radius $r = 7$ (with $\phi = 2$ within the droplet) on every point of a triangular lattice (spacing $\ell_t = 8$) with some noise. We set the number of cells in each row to be $\sqrt{N}$. Thus, the dimensions of the full lattice are $160 \times 138$ ($96\times83$) for $N = 100$ ($N = 36$). We then allow these cells to relax and grow (with $\text{Pe} = 0$) for $10^4$ timesteps such that the system becomes confluent. \mc{Confluence is achieved because the cell target area $\pi R^2$ is larger than the area available to each cell, which renders the force associated with the third term in Eq.~(\ref{eqn:free_energy}) qualitatively equivalent to a negative pressure. To ensure near-constant cell area and confluent conditions at all times, we use  \mc{$\lambda \geq 3000\,K$}}. Next, we switch on motility, \mc{by varying $v_0$} and evolve the cells for $10^6$ timesteps. In the main production runs, we sample the system every $10^3$ timesteps for $2\times10^7$ timesteps

\begin{figure*}[ht]
  \includegraphics[width=\textwidth]{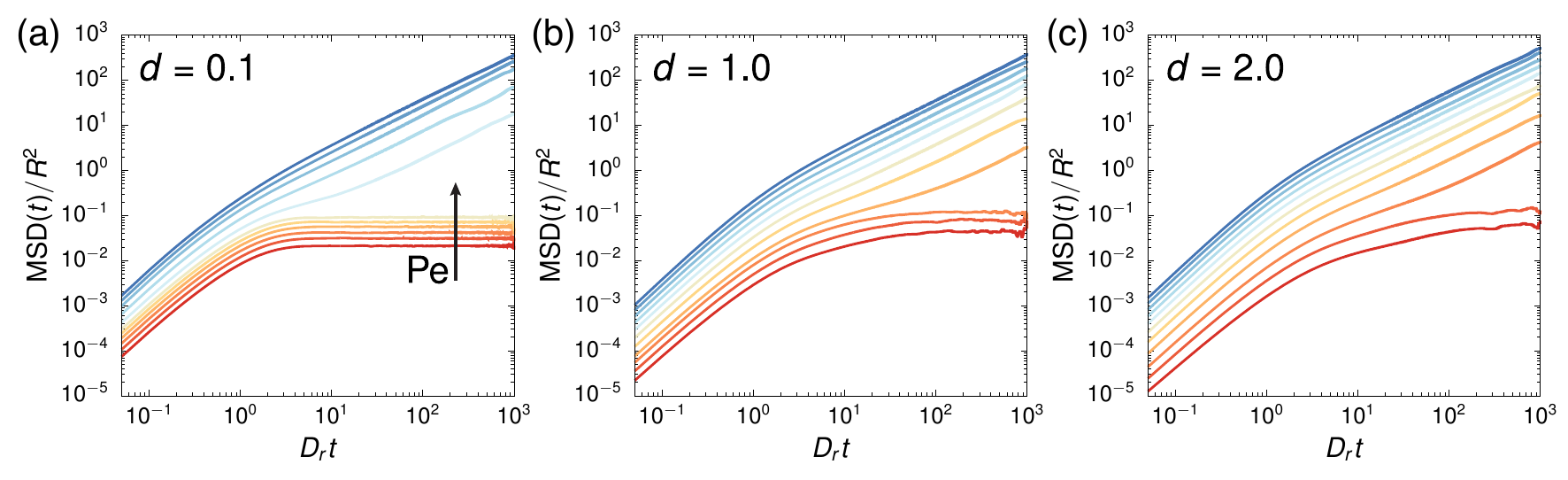}
  \caption{Mean square displacement (MSD) for (a) $d = 0.1$, (b) $1.0$, and (c) $2.0$. In each plot, we show MSD curves for $\text{Pe}$ values ranging from $1.0$ to $3.0$ in increments of $0.2$. Each curve is averaged over three simulation runs.}
  \label{fig:msd}
\end{figure*}

\begin{figure}[ht]
  \includegraphics[width=0.5\textwidth]{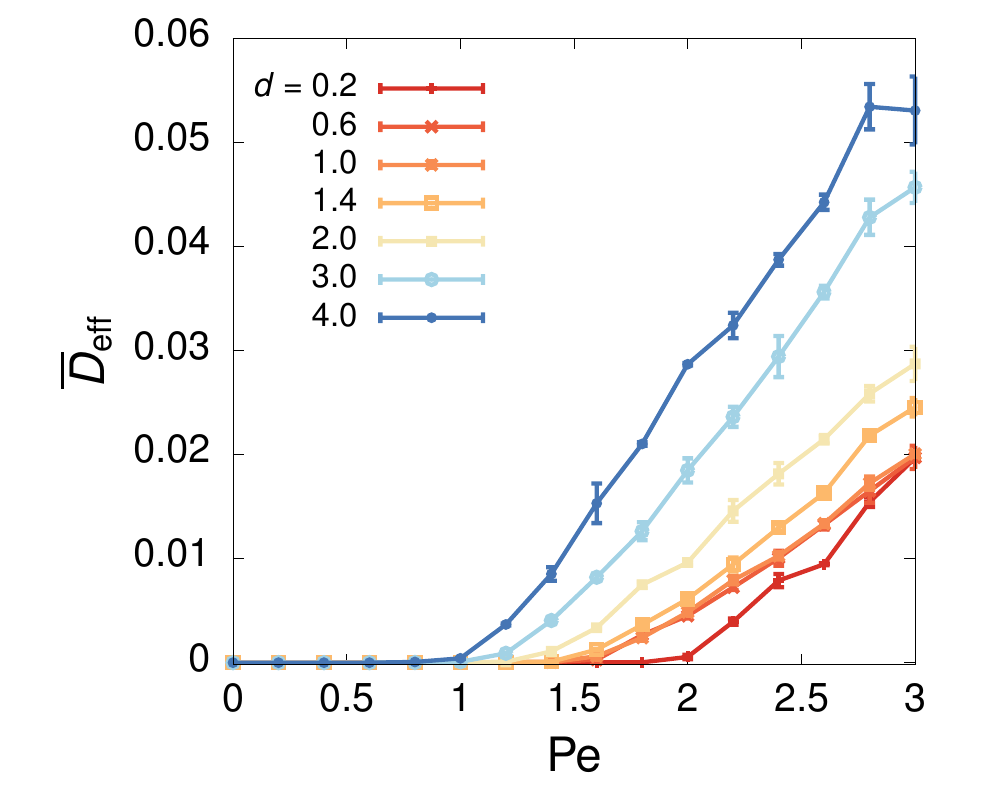}
  \caption{Normalized effective diffusivity $\overline{D}_{\text{eff}}$ as a function of motility $\text{Pe}$ for a range of deformability values. Each point is sampled from three simulation runs.}
  \label{fig:deff}
\end{figure}
\begin{figure*}
  \includegraphics[width=0.9\textwidth]{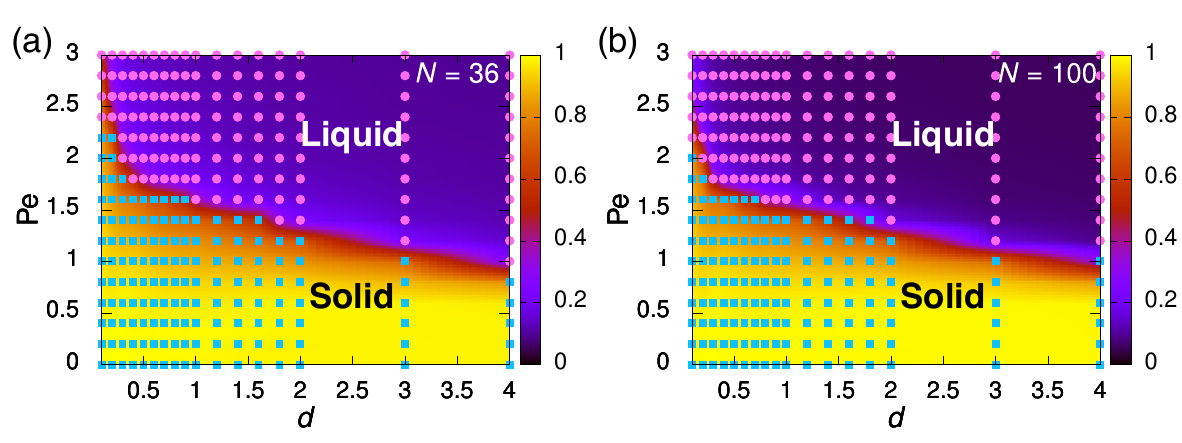}
  \caption{Quantifying the solid-liquid transition based on the system's effective diffusivity $\overline{D}_{\text{eff}}$ and its global bond-orientational order $\abs{\Psi_6}$, for (a) $N = 36$ and (b) $N = 100$. The points on these plots indicate the locations that we have explored in the parameter space. The results are averaged over three simulation runs. At these points, the system is labeled as a liquid (pink circles) if $\overline{D}_{\text{eff}} > 0.0005$ or a solid (blue squares) otherwise. The heat maps in the background are constructed from interpolating the measured $\abs{\Psi_6}$ values at these points.}
  \label{fig:phase_diagram_hexatic}
  \vspace{1cm}
\end{figure*}
\noindent (or $D_rt = 1000$). We explore the parameter space $d = 0.1$--$4.0$ and $\text{Pe} = 0.0$--$3.0$ to locate the solid and liquid regimes of the system.

\section{DYNAMICAL OBSERVABLES}

\subsection{Mean Square Displacement (MSD)}
To pinpoint the solid-liquid transition, we first examine dynamical observables. We compute the mean square displacement (MSD) of the cells as
\begin{align}
\text{MSD}(t) = \avg{\frac{1}{N}\sum_{i=1}^{N}({\vec{R}'}_i^{\text{cm}}(t+\tau)-{\vec{R}'}_i^{\text{cm}}(\tau))^2}_{\tau},
\end{align}
where ${\vec{R}'}_i^{\text{cm}}$ is the center of mass of the $i$th cell in the rest frame of the full monolayer. Fig.~S\ref{fig:msd} shows the MSD curves for several  values of deformability. They suggest that increasing deformability $d$ facilitates melting, with lower motility required. Further, the region in which the MSD curve plateaus shrinks with increasing $d$, indicating that cells can squeeze past each other more easily as cells become more deformable.

\subsection{Effective Diffusivity}
From the MSD data, we determine the dynamical arrest of the system by calculating a normalized effective diffusivity
\begin{align}
\overline{D}_{\text{eff}} = \lim_{t\to\infty}\frac{\text{MSD}(t)}{4D_0t},
\end{align}
where $D_0 = v_0^2/(2D_r)$ is the diffusivity of an isolated active Brownian particle undergoing rotational diffusion. In practice, we measure $\overline{D}_{\text{eff}}$ by performing linear fits of the MSD curves at late times ($5\times10^6$ to $1.5\times10^7$ timesteps, or $D_rt = 250$ to $750$) and using the slope of the fit to estimate the diffusivity. Fig.~S\ref{fig:deff} reports the measured $\overline{D}_{\text{eff}}$ for a range of deformability values as a function of motility. The plot suggests that the system fluidizes (i.e., $\overline{D}_{\text{eff}} > 0$) at lower motility as they become more deformable.

We identify the system as a liquid if $\overline{D}_{\text{eff}} > 0.0005$ and a solid otherwise. This threshold is chosen to match the results from structural observables (see below). Fig.~S\ref{fig:phase_diagram_hexatic} displays a phase diagram constructed based on this criterion (and also the structural observables) for $N = 36$ and $100$. The transition boundary is similar in both system sizes for $d > 1$, but it occurs at a slightly lower motility value as $N$ increases for $d < 1$. 

\section{STRUCTURAL OBSERVABLES}
\subsection{Bond-Orientational Order Parameter}
We measure structural observables as an alternative way to quantify the melting transition. Since we initialize the cells in a regular hexagonal arrangement, these cells have near perfect $6$-fold coordination if the system remains a solid. On the other hand, when the system melts, cells exchange positions with one another, and their nearest neighbors are not, in general, arranged in a hexagonal manner. A suitable observable which captures the change in orientational order is the \textit{local} bond-orientational order parameter $\Psi_{6,j}$, which is defined for each cell (say cell $j$) as
\begin{align}
\Psi_{6,j}(t) = \frac{1}{N_{\text{nn},j}}\sum_{k\in\text{nn}}\exp\left[i6\theta_{jk}(t)\right],
\end{align}
where the sum is over the nearest neighbors of the cell, $N_{\text{nn},j}$ is its number of nearest neighbors, and $\theta_{jk}$ is the angle between the line connecting the center of mass of cell $j$ and $k$ and a reference axis (taken to be the $x$ axis here). We use Delaunay triangulation to determine the nearest neighbors of each cell when computing this observable. The \textit{global} bond-orientational order parameter $\Psi_6(t)$ is defined as the average of $\Psi_{6,j}(t)$ over all cells. Note that $\abs{\Psi_6} \simeq 1$ when cells have nearest neighbors close to perfect hexagonal arrangement (i.e., in the solid state), whereas $\abs{\Psi_6} \simeq 0$ when local orientational order is lost (i.e., in the liquid state). In Fig.~S\ref{fig:phase_diagram_hexatic}, we display the heat maps of $\abs{\Psi_6}$ underneath the phase diagram constructed based on the dynamical criterion on $\overline{D}_{\text{eff}}$. Notably, the region where $\abs{\Psi_6} \simeq 1$ is in agreement with the points classified as a solid, whereas $\abs{\Psi_6} \simeq 0$ maps to the points labeled as a liquid.

\begin{figure*}[ht]
  \centering
  \includegraphics[width=\textwidth]{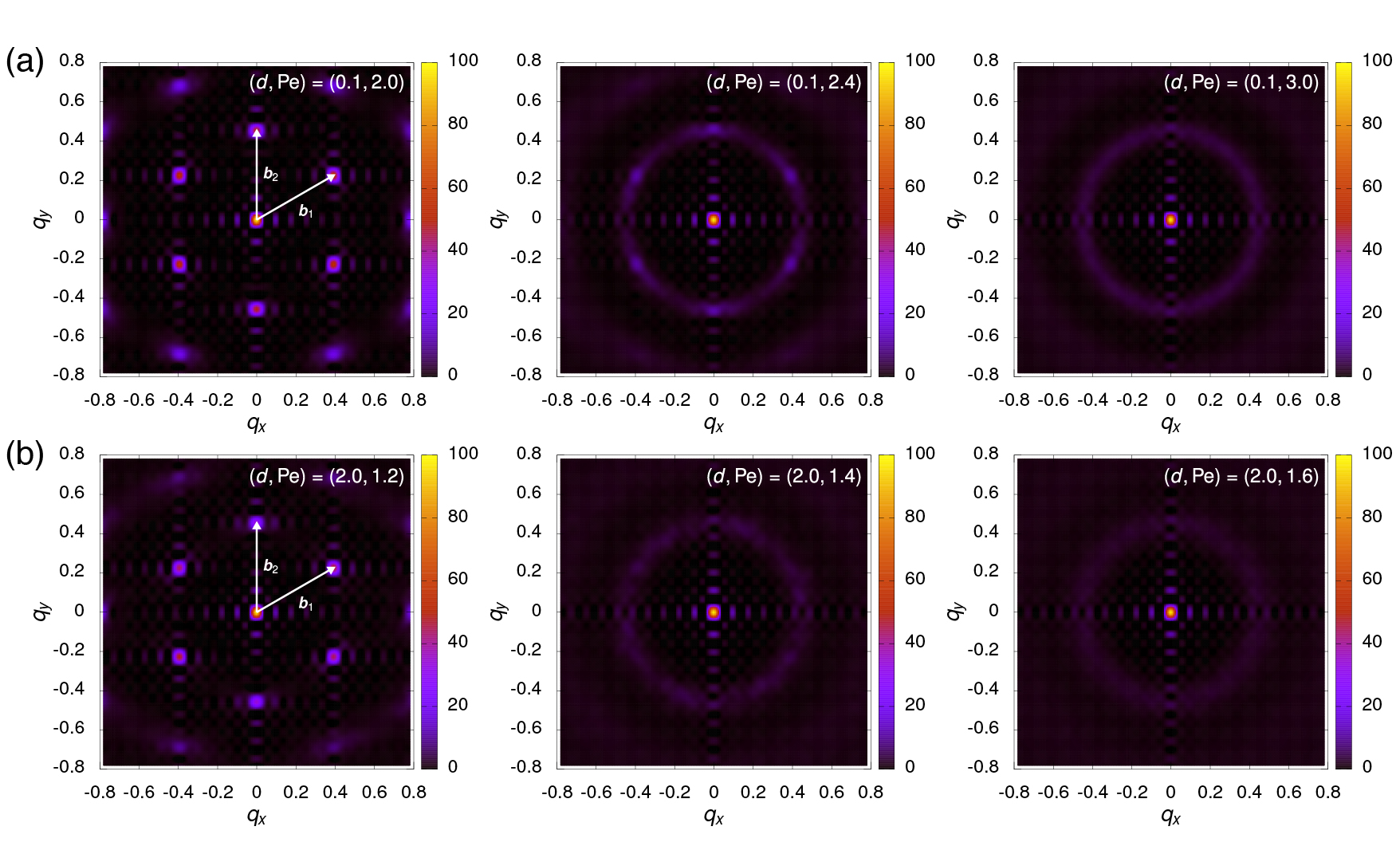}
  \caption{Structure factor $S(\vec{q})$ for (a) $d = 0.1$ and (b) $d = 2.0$ at three different $\text{Pe}$ values across the solid-liquid transition boundary. Each plot shows data from a single simulation. The white arrows indicate the reciprocal lattice vectors, $\vec{b}_1$ and $\vec{b}_2$. \mc{Note that $S(\vec{q})$ can vary between solid- and liquid-like at $(d,\text{Pe}) = (0.1,2.4)$, which is within the intermittent region.}}
  \label{fig:structure_factor}
\end{figure*}

\subsection{Structure Factor}
In addition to $\Psi_6$, we analyze the structure factor $S(\vec{q})$ of the system, which is given by
\begin{align}
S(\vec{q}) = \avg{\frac{1}{N}\sum_{i=1}^N\sum_{j=1}^{N}e^{i\vec{q}\cdot(\vec{R}_i^{\text{cm}}-\vec{R}_j^{\text{cm}})}}.
\end{align}
In the ideal crystalline state, cells are arranged in a regular triangular lattice with spacing $\ell_t$ between lattice points. The real space lattice can be defined by two lattice vectors forming a unit cell:
\begin{align}
\vec{a}_1 = \left(\ell_t,0\right) \qquad \vec{a}_2 = \frac{\ell_t}{2}\left(-1,\sqrt{3}\right).
\end{align}
The corresponding reciprocal lattice vectors are
\begin{align}
\vec{b}_1 = \frac{2\pi}{\ell_t}\left(1,\frac{1}{\sqrt{3}}\right) \qquad \vec{b}_2 = \frac{2\pi}{\ell_t}\left(0,\frac{2}{\sqrt{3}}\right).
\end{align}
Fig.~S\ref{fig:structure_factor} shows $S(\vec{q})$ at several points near the transition boundary for both low and high deformability values. The data are consistent with those from other observables. In particular, for points that are labeled as a solid, $S(\vec{q})$ has maxima at the reciprocal lattice points, indicating that the system exhibits translational order. On the other hand, for points in the fluid regime, these maxima fade away and $S(\vec{q})$ becomes isotropic (as demonstrated by the formation of a ring).

\begin{figure*}[ht]
  \includegraphics[width=0.9\textwidth]{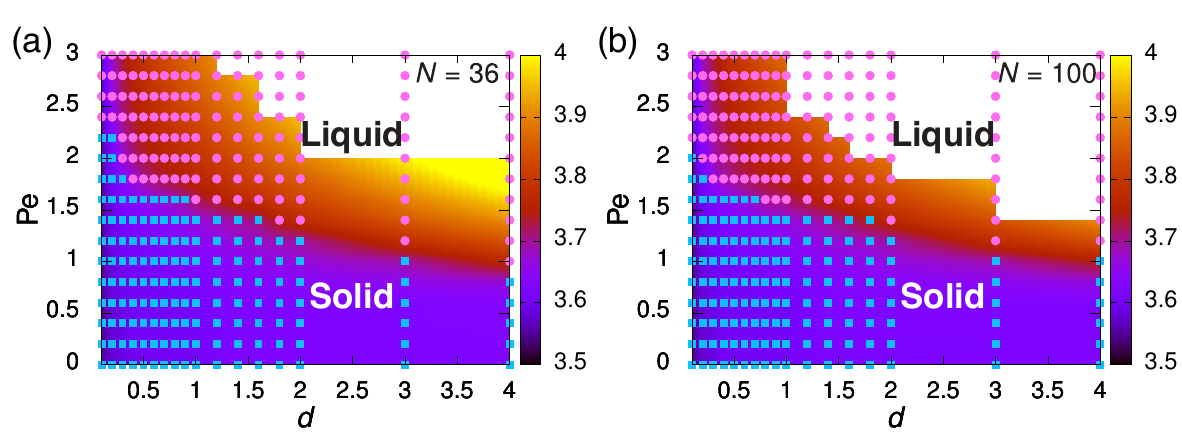}
  \mc{\caption{Heat maps showing the shape index $q$ of cells within the explored parameter space for (a) $N = 36$ and (b) $N = 100$. As in Fig.~S\ref{fig:phase_diagram_hexatic}, we superpose the maps with the sampled data points color-coded as liquid (pink circles) or solid (blue squares) based on the effective diffusivity $\overline{D}_{\text{eff}}$. Data are not available at high deformability $d$ and motility $\text{Pe}$ as cells deform substantially in this regime and may not have well-defined, closed contours.}
  \label{fig:phase_diagram_shape_index}}
\end{figure*}

\begin{figure*}[ht]
  \includegraphics[width=0.9\textwidth]{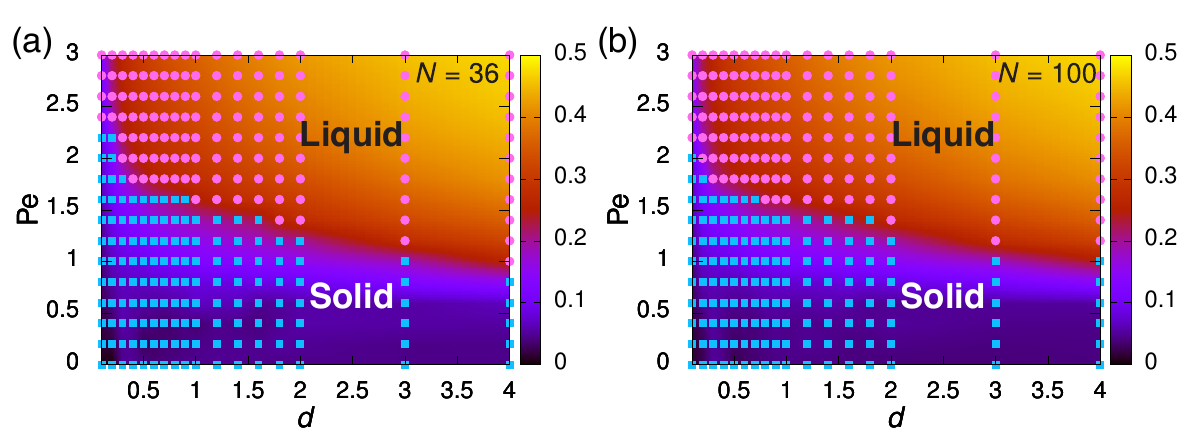}
  \mc{\caption{Heat maps showing the eccentricity $\mathcal{E}$ of cells within the explored parameter space for (a) $N = 36$ and (b) $N = 100$. As above, data points are color-coded as liquid (pink circles) or solid (blue squares) based on the effective diffusivity $\overline{D}_{\text{eff}}$.}
  \label{fig:phase_diagram_eccent}}
\end{figure*}

\mc{
\subsection{Cell Shape Index and Eccentricity}
For completeness, we present two other structural observables that are of interest when modeling cell monolayers. First, in line with vertex and Voronoi model studies, we examine the cell shape index
\begin{align}
q = \avg{\frac{1}{N}\sum_{i=1}^{N} \frac{P_i}{\sqrt{A_i}}},
\end{align}
where $P_i$ and $A_i$ are the perimeter and area, respectively, of the contour ($\phi_i = 1$) of cell $i$. Note that $q\mc{=2\sqrt{\pi}} \approx 3.54$ for a circle and $q = 2^{3/2}\, 3^{1/4} \approx 3.72$ for a regular hexagon. Second, as the cells in our model are deformable, we quantify their degree of elongation by measuring their eccentricity
\begin{align}
\mathcal{E} = \avg{\frac{1}{N}\sum_{i=1}^{N}\frac{\abs{\lambda_{i,+}-\lambda_{i,-}}}{\lambda_{i,+}+\lambda_{i,-}}},
\end{align}
where $\lambda_{i,+}$ and $\lambda_{i,-}$ are the eigenvalues of the gyration tensor $G$ of cell $i$, whose components are given by
\begin{align}
  G_{i,\alpha\beta} = \frac{1}{A_i}\int_{\Omega_{i}} d^2\vec{r}\,\left(r_{\alpha}-R_{i,\alpha}^{\text{cm}}\right)\left(r_{\beta}-R_{i,\beta}^{\text{cm}}\right),
\end{align}
where $\Omega_i$ is the region in which $\phi_i > 1$.  With this definition, $\mathcal{E} \approx 0$ when cells are circular and $\mathcal{E} \approx 1$ when they are rod-like. Fig.~S\ref{fig:phase_diagram_shape_index} and Fig.~S\ref{fig:phase_diagram_eccent} show heat maps of the shape index and eccentricity, respectively, for $N = 36$ and $100$. The maps indicate these two observables are correlated with one another. This is reasonable given all cells have similar areas under the strong \mc{area} constraint \mc{used in our simulations} (see further below), and cell elongation is likely to lengthen a cell's perimeter without changing much its area, thus resulting in a higher shape index. Although these observables capture the solid-liquid transition well at high deformability ($d > 1$), they are less effective in quantifying the transition at low deformability ($d < 1$). This is because cells can exchange neighbors via overlapping apart}

\mc{\noindent from deforming their boundaries when the system fluidizes in this regime. 
}

\begin{figure*}[ht]
  \centering
  \includegraphics[width=\textwidth]{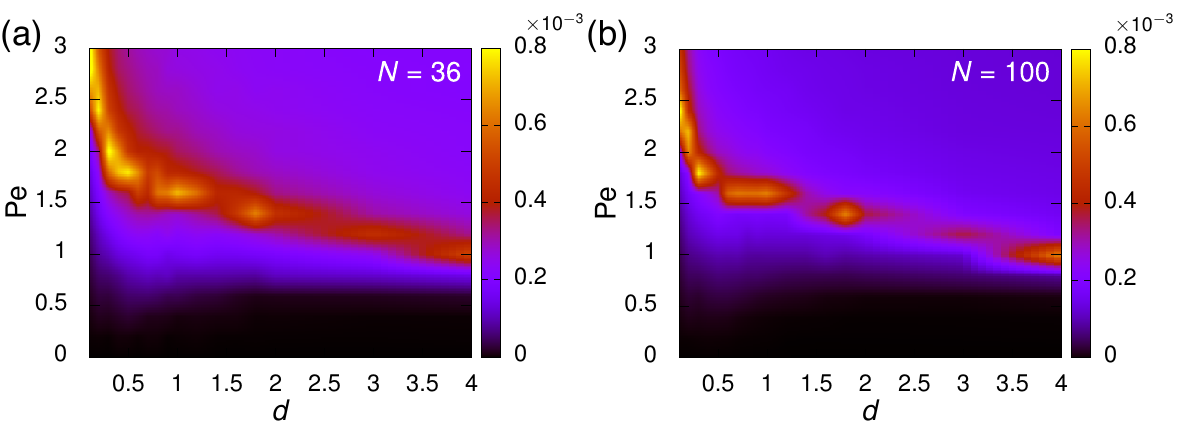}
  \caption{Heat maps reporting the standard error of $\abs{\Psi_6}$ within the explored parameter space for (a) $N = 36$ and (b) $N = 100$.}
  \label{fig:phase_diagram_hex_var}
  \vspace{1cm}
  \centering
  \includegraphics[width=0.9\textwidth]{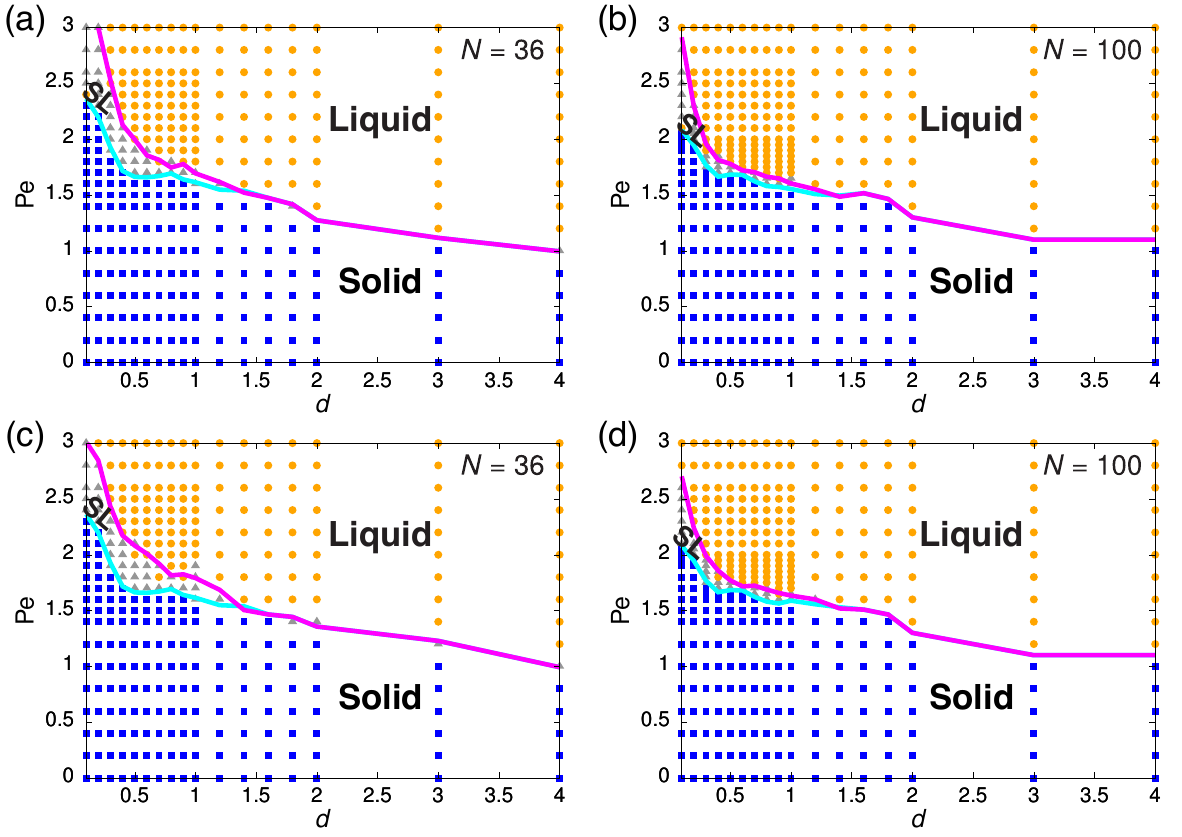}
  \caption{Phase diagrams constructed from analyzing (a,b) the standard error of $\abs{\Psi_6}$ and (c,d) the binarized signal for the fraction of cells with defects $\avg{\abs{\Delta N_{\text{nn}}}}$, for both $N = 36$ and $100$. The liquid (orange circles) and solid phases (blue squares) are identified based on the criterion on the effective diffusivity, $\overline{D}_{\text{eff}} > 0.0005$. In (a,b), the intermittent region (labeled as SL; gray triangles) is located where the standard error of $\abs{\Psi_6}$ and the value of $\overline{D}_{\text{eff}}$ are above $0.0005$. In (c,d), this region is identified as where the binarized defect signal shows more than two jumps and that $\overline{D}_{\text{eff}} > 0.0005$. The cyan (magenta) line indicates the lower (upper) bound of the intermittent region and is computed from interpolating the boundaries separating the three regimes. The phase diagram shown in Fig.~2 in the main paper is based on the boundaries drawn in (b).}
  \label{fig:phase_diagram_intermit}
\end{figure*}

\section{INTERMITTENT STATE}
At low deformability ($d < 1$) there is an intermittent region near the transition boundary in which the system experiences episodes of fluidization and freezing events. This is demonstrated in the time series of the bond-orientational order $\abs{\Psi_6}$ (Fig.~3). We employ two different methods to locate this region. 

First, we examine the standard error of $\abs{\Psi_6}$, which measures the fluctuations between solid and liquid states. We calculate this quantity using the time series data from three simulation runs for each point sampled in the parameter space. Fig.~S\ref{fig:phase_diagram_hex_var} reports heat maps of this standard error. One can see that the region with large standard error shrinks as deformability increases. We associate a state to be within the intermittent regime if both the standard error of $\abs{\Psi_6}$ and the effective diffusivity $\overline{D}_{\text{eff}}$ are above $0.0005$. The resulting phase diagrams are displayed in Fig.~S\ref{fig:phase_diagram_intermit}a,b.

\begin{figure}[ht]
  \centering
  \includegraphics[width=0.5\textwidth]{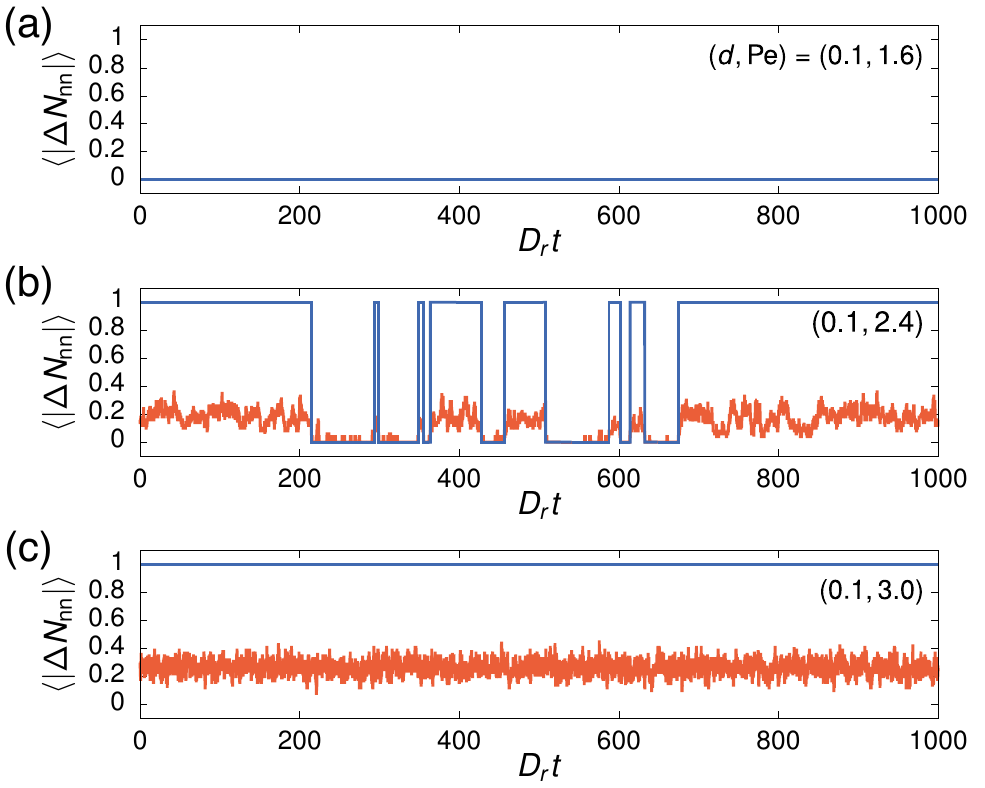}
  \caption{Time series for the fraction of cells with defects $\avg{\abs{\Delta N_{\text{nn}}}}$ in the system for $d = 0.1$ when it is in the (a) solid ($\text{Pe = 1.6}$), (b) intermittent ($\text{Pe = 2.4}$), and (c) liquid ($\text{Pe = 3.0}$) regime (from a single simulation run). The orange curves show the measured $\avg{\abs{\Delta N_{\text{nn}}}}$ over time, whereas the blue curves are the binarized version of the signal with zero for solid and unity for liquid.}
  \label{fig:defects}
\end{figure}

Second, we note that intermittence is not only associated with large fluctuations, but also requires that such fluctuations be correlated over finite times, hence the system spends a finite fraction of time in each state.  To quantify time-correlations, we examine the time persistence in fluctuations in the number of disclinations. We define a ``topological charge'' for each cell (say cell $i$) as its deviation from a $6$-fold coordination,
\begin{align}
\Delta N_{\text{nn},i}(t) = N_{\text{nn},i}(t)-6,
\end{align}
where $N_{\text{nn},i}$ is the number of nearest neighbors of the cell. We associate cells with $\Delta N_{\text{nn},i} < 0$ as negative defects, whereas those with $\Delta N_{\text{nn},i} > 0$ as positive defects. We estimate the fraction of cells with defects as
\begin{align}
\avg{\abs{\Delta N_{\text{nn}}}}(t) = \frac{1}{N}\sum_{i=1}^{N}\abs{\Delta N_{\text{nn},i}(t)}.
\end{align}
Since there are few defects in the solid regime, we can identify the system as solid-like when the time series of $\avg{\abs{\Delta N_{\text{nn}}}}$ is below $5\%$, and liquid-like when it goes above this threshold. This allows us to binarize the defect  time series, with zero for solid and unity for liquid. To ensure that the solid and liquid states are persistent in time, we smooth the data by removing any jumps in the signal that lasts less than $5D_r^{-1}$. We label states as intermittent if there are at least two jumps in the binarized signal, and if additionally the system's effective diffusivity $\overline{D}_{\text{eff}}$ is above $0.0005$. The latter condition ensures that the system is actually fluidized when there are many defects present. Fig.~S\ref{fig:defects} reports examples of the time series of $\avg{\abs{\Delta N_{\text{nn}}}}$ and its corresponding binarized signal for a solid, a liquid, and an intermittent state. Fig.~S\ref{fig:phase_diagram_intermit}c,d report phase diagrams with the intermittent region identified based on the algorithm discussed. The parameter space spanned by this region is consistent with that estimated based on the standard error of $\abs{\Psi_6}$.\mc{ Interestingly, the phase diagrams in Fig.~S\ref{fig:phase_diagram_intermit} show that the intermittent region shrinks in size as the system becomes larger. This remains the case when we perform additional simulations for $N = 400$ (Fig.~S\ref{fig:intermit_region}). Such reduction in intermittence is expected in the context of finite size effects for a first-order-like transition.}

\mc{

\begin{figure*}[ht]
  \centering
  \includegraphics[width=\textwidth]{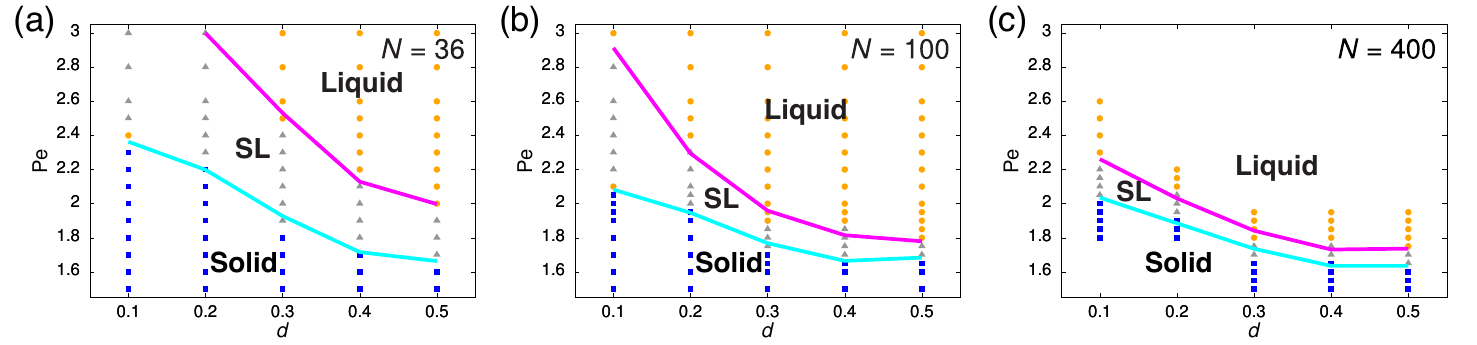}
  \mc{\caption{Enlarged view of the intermittent region (SL) at low deformability for (a) $N = 36$, (b) $N = 100$, and (c) $N = 400$, as computed based on the standard error of $\abs{\Psi_6}$ and the effective diffusivity $\overline{D}_{\text{eff}}$ (see Fig.~S\ref{fig:phase_diagram_intermit}a,b). The region shrinks with $N$.}
  \label{fig:intermit_region}}
\end{figure*}
  
\begin{figure}[ht]
  \centering
  \includegraphics[width=0.5\textwidth]{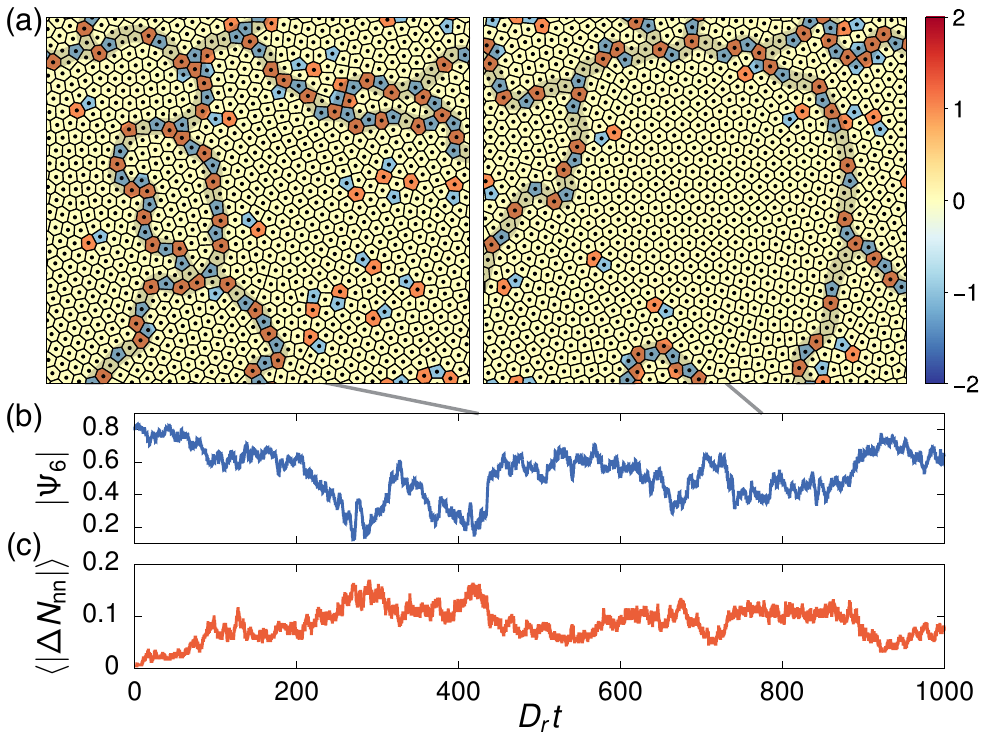}
  \mc{\caption{Spatial clustering of topological defects in the intermittent region $(d,\text{Pe}) = (0.1,2.05)$ with $N = 900$. (a) Simulation snapshots showing that defects tend to form long chains and clusters. The light gray shading highlights potential grain boundaries within the system. (b)-(c) Time series of $\abs{\Psi_6}$ and $\avg{\abs{\Delta N_{\text{nn}}}}$ for this simulation run.}
  \label{fig:defects_percolation}}
\end{figure}
  
A recent simulation study on 2D passive and active repulsive disk systems shows that topological defects can form clusters and tend to percolate across the system near the melting transition~\cite{Digregorio2019}. We examine here qualitatively the spatial distribution of defects within the intermittent region using a larger system $N = 900$. In line with their observations, we find defects often form lines and grain boundaries (despite missing a few defects occasionally) which seem to extend across the system (Fig.~S\ref{fig:defects_percolation}). This is also observed in smaller systems (see snapshots in Fig.~3). Nevertheless, it is difficult to draw further conclusions given the relatively limited number of cells in our work. It would, therefore, be of interest in the future to analyze quantitatively the nature of defect percolation in this model and its relation to the intermittent behavior with a much larger system. 
}

\mc {
\section{FINITE SIZE SCALING ANALYSIS}
The existence of an intermittent state at low deformability prompts us to further investigate the nature of the transition  in this regime. To this end, we perform a finite size scaling analysis at $d = 0.1$ spanning systems of size $N = 36, 100, 256, 400$, and $900$. Treating $\abs{\Psi_6}$ as the order parameter, we consider the scaling ansatz
\begin{align}
  \abs{\Psi_6} = N^{\zeta}f(pN^{\nu}),
\end{align}
where $\nu$ and $\zeta$ are scaling exponents to be determined, and $p = \text{Pe}/\text{Pe}^{*}-1$ with $\text{Pe}^{*}$ the critical P\'eclet value as $N\to\infty$. We estimate $\text{Pe}^{*}$, $\nu$, and $\zeta$ numerically by means of the procedure outlined in Ref.~\cite{Bhattacharjee2001}. Specifically, we compute a residual function $\mathcal{R}$ that measures the pairwise differences between data sets from different $N$ when scaled by a given set of $(\text{Pe}^{*},\nu,\zeta)$:
\begin{align}
\mathcal{R}(\text{Pe}^{*},\nu,\zeta) = \frac{1}{\mathcal{N}}\sum_{i}\sum_{j\neq i}\sum_{k,\text{over}}\abs{\abs{\Psi_6}_{jk}N_{j}^{-\zeta}-\mathcal{L}_i(p_{jk}N_{j}^{\nu})},
\end{align}
where the first two sums are performed over all possible pairs of data sets $i,j$, and the innermost sum is carried out over the data points $k$ of set $j$ that are within the rescaled domain of set $i$. $\mathcal{L}_i(x)$ is an interpolating function based on the values of set $i$ (we use a four-point Lagrange interpolation polynomial), and $\mathcal{N}$ is the number of data points compared in the sums. Minimizing $\mathcal{R}$ using the Nelder-Mead algorithm~\cite{NumericalRecipes1992}, we find the parameters which best collapse our data sets are $\text{Pe}^{*} = 1.967(7)$, $\nu = 0.75(2)$, and $\zeta = -0.044(3)$ (Fig.~S\ref{fig:hexatic_scaling}). The errors are estimated based on $1\%$ deviation from the minimum residual $\mathcal{R}_0$. The finding $\zeta \approx 0$ indicates that $\abs{\Psi_6}$ is likely to be discontinuous at the transition line; as $N\to\infty$, $|\Psi_6|\to 0$ for  $p>0$ but remains finite at $p=0$. This is consistent with the observation of intermittence near the phase boundary and that the transition is first-order-like at low deformability.

\begin{figure}[ht]
  \centering
  \includegraphics[width=0.5\textwidth]{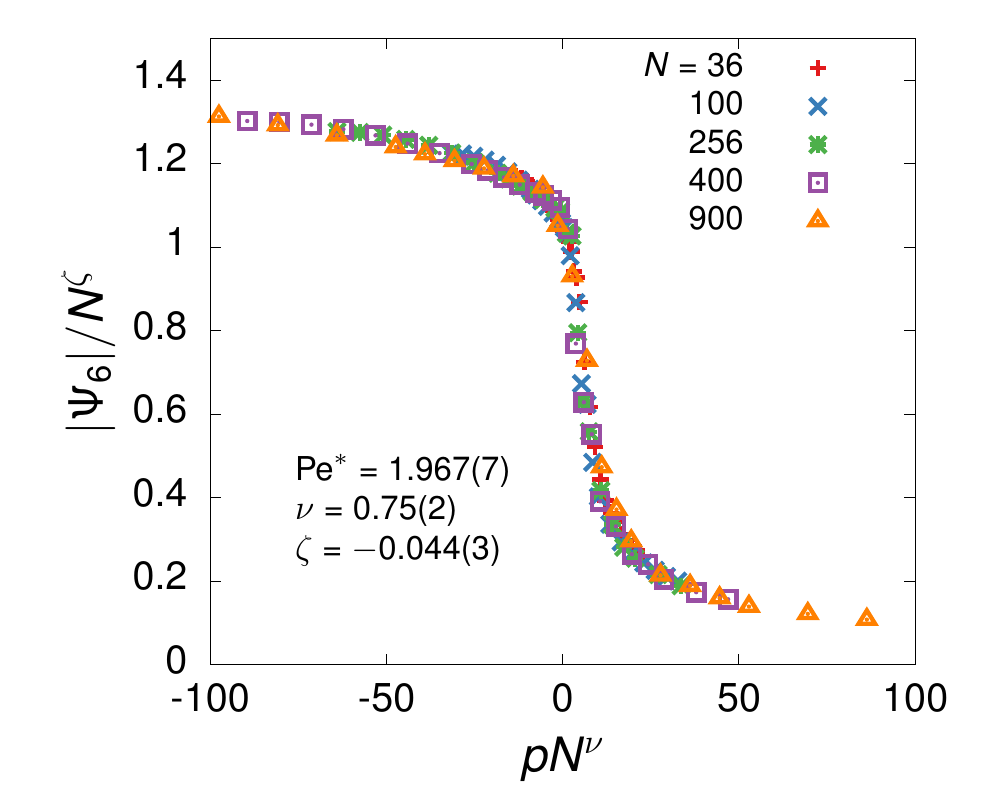}
  \mc{\caption{Collapse of the data for $\abs{\Psi_6}$ from different system sizes $N$ based on our estimate of the scaling parameters ($\text{Pe}^{*},\nu,\zeta)$, which are obtained from minimizing the residual $\mathcal{R}$.}
  \label{fig:hexatic_scaling}}
\end{figure}
}

\section{LOCAL OVERLAP}

\begin{figure*}[ht]
  \centering
  \includegraphics[width=\textwidth]{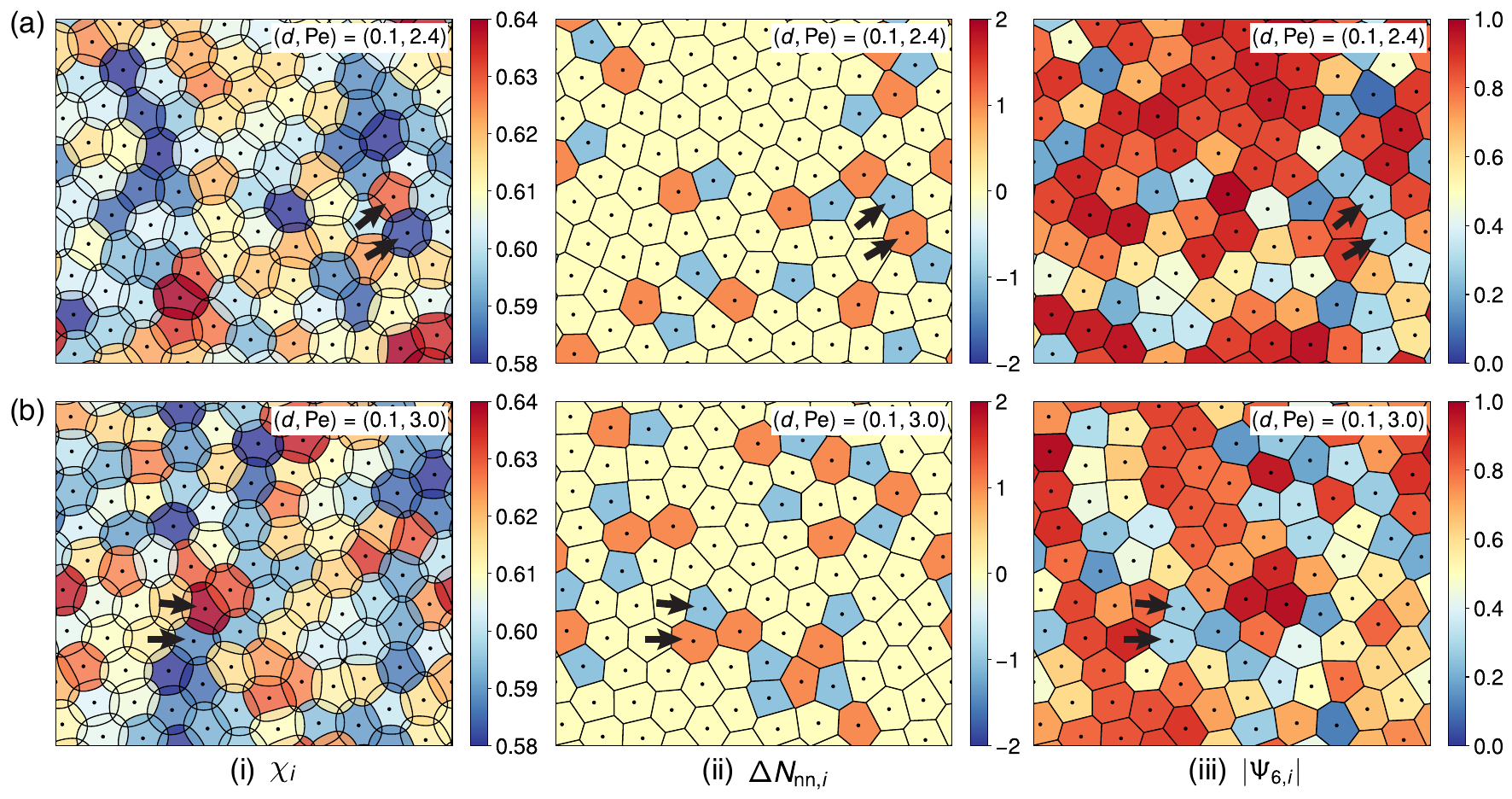}
  \caption{Simulation snapshots of the system for (a) $(d,\text{Pe}) = (0.1,2.4)$ and (b) $(0.1,3.0)$, showing the correlation between (i) the degree of local overlap $\chi_i$, (ii) the topological charge $\Delta N_{\text{nn},i}$, and (iii) the local bond-orientational order $\abs{\Psi_{6,i}}$. In (ii) and (iii), we plot a Voronoi deconstruction of the system to aid visualization. Black arrows highlight a positive (a cell with $7$-fold disclination) and a negative defect (a cell with $5$-fold disclination). Negative defects have a higher degree of local overlap than positive ones. Moreover, in the Voronoi representation, the polygons associated with negative defects have smaller areas than those associated with positive defects.}
  \label{fig:overlap_defect_hexatic}
\end{figure*}

\begin{figure*}[ht]
  \centering
  \includegraphics[width=0.8\textwidth]{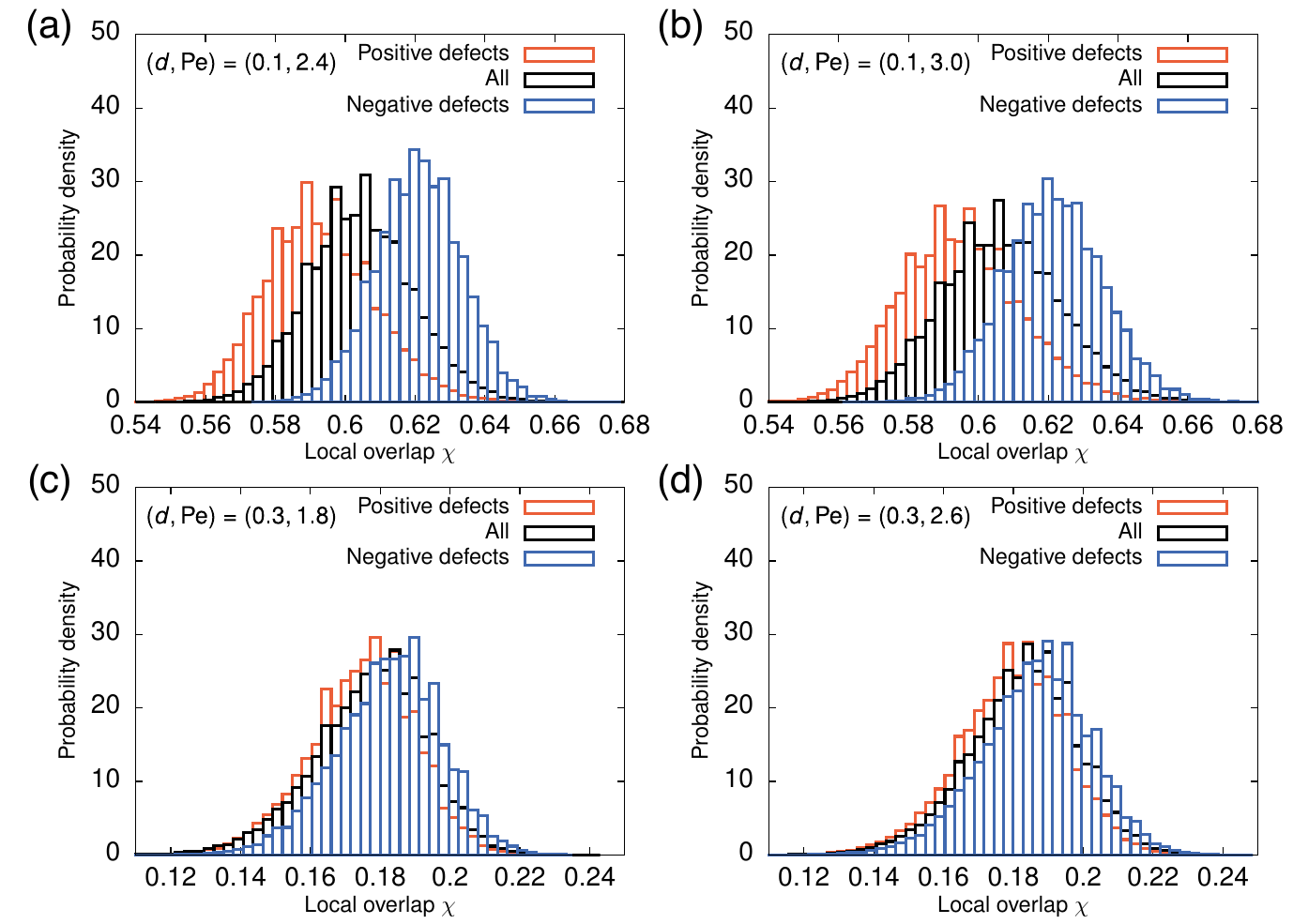}
  \caption{Probability density of the local overlap $\chi$ for negative defects, positive defects, and all cells for four different parameter sets: (a) $(d,\text{Pe}) = (0.1,2.4)$, (b) $(0.1,3.0)$, (c) $(0.3,1.8)$, and (d) $(0.3,2.6)$. In each case, the distribution is constructed from sampling time frames from two separate simulation runs in which defects exist. Note that negative defects tend to have a higher degree of overlap than positive defects.}
  \label{fig:overlap_distribution}
\end{figure*}

To analyze the correlation between defects and cell overlap, we define a local overlap field $\chi_i(\vec{r})$ for each cell (say cell $i$) as
\begin{align}
\chi_i(\vec{r}) = \sum_{j=1}^{N}H(\phi_i(\vec{r})-1)H(\phi_j(\vec{r})-1),
\end{align}
where $H(x)$ is the Heaviside step function (i.e., we only consider the region within a cell where $\phi > 1$). This field is unity at sites where the cell overlaps with another cell and zero otherwise (it can exceed unity when there are multiple pairwise overlaps). We estimate the area fraction of a cell overlapped by other cells as
\begin{align}
\chi_i = \frac{1}{A_i}\int_{\Omega_i}\fd^2\vec{r}\,\chi_i(\vec{r}),
\end{align}
where $\Omega_i$ is the region where $\phi_i > 1$ and $A_i$ is the area of this region. Fig.~S\ref{fig:overlap_defect_hexatic} presents simulation snapshots of the system highlighting the degree of local overlap $\chi_i$ of each cell, its local bond-orientational order $\Psi_{6,i}$, and the location of defects. Not surprisingly, cells with defects are those with a lower orientational order, as they do not have a $6$-fold coordination. More important, a careful inspection reveals these defects correlate with the degree of overlap, with negative defects showing more overlap than positive defects. To quantify this observation, we look at time frames in which defects exist when the system is in the intermittent or liquid regime, and we construct distributions of the local overlap $\chi$ for  defects and for all cells (Fig.~S\ref{fig:overlap_distribution}). These distributions reinforce the visual impression that negative defects indeed have a higher degree of overlap than the average population, whereas positive defects have a lower degree of overlap. The deviation between these distributions is also significant as quantified by the non-parametric Kolmogorov-Smirnov (KS) two-sample test (Table~S\ref{table:overlap_ks_test}).

\begin{figure*}[ht]
  \centering
  \includegraphics[width=0.8\textwidth]{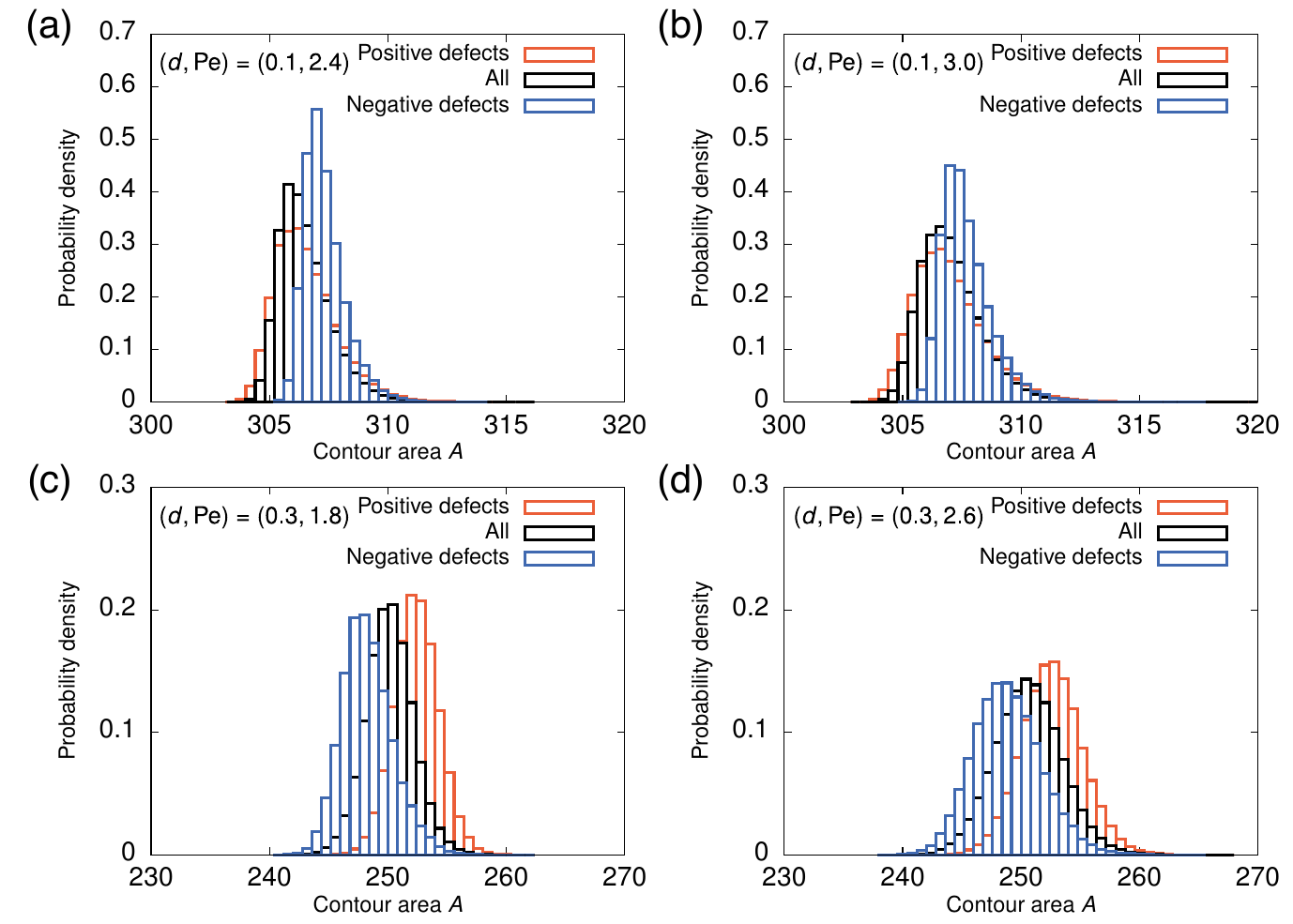}
  \mc{\caption{Probability density of the contour area $A$ of a cell for negative defects, positive defects, and all cells for the same parameter sets as in Fig.~S\ref{fig:overlap_distribution}.}
    \label{fig:area_distribution}}
  \vspace{0.5cm}
  \centering
  \includegraphics[width=0.8\textwidth]{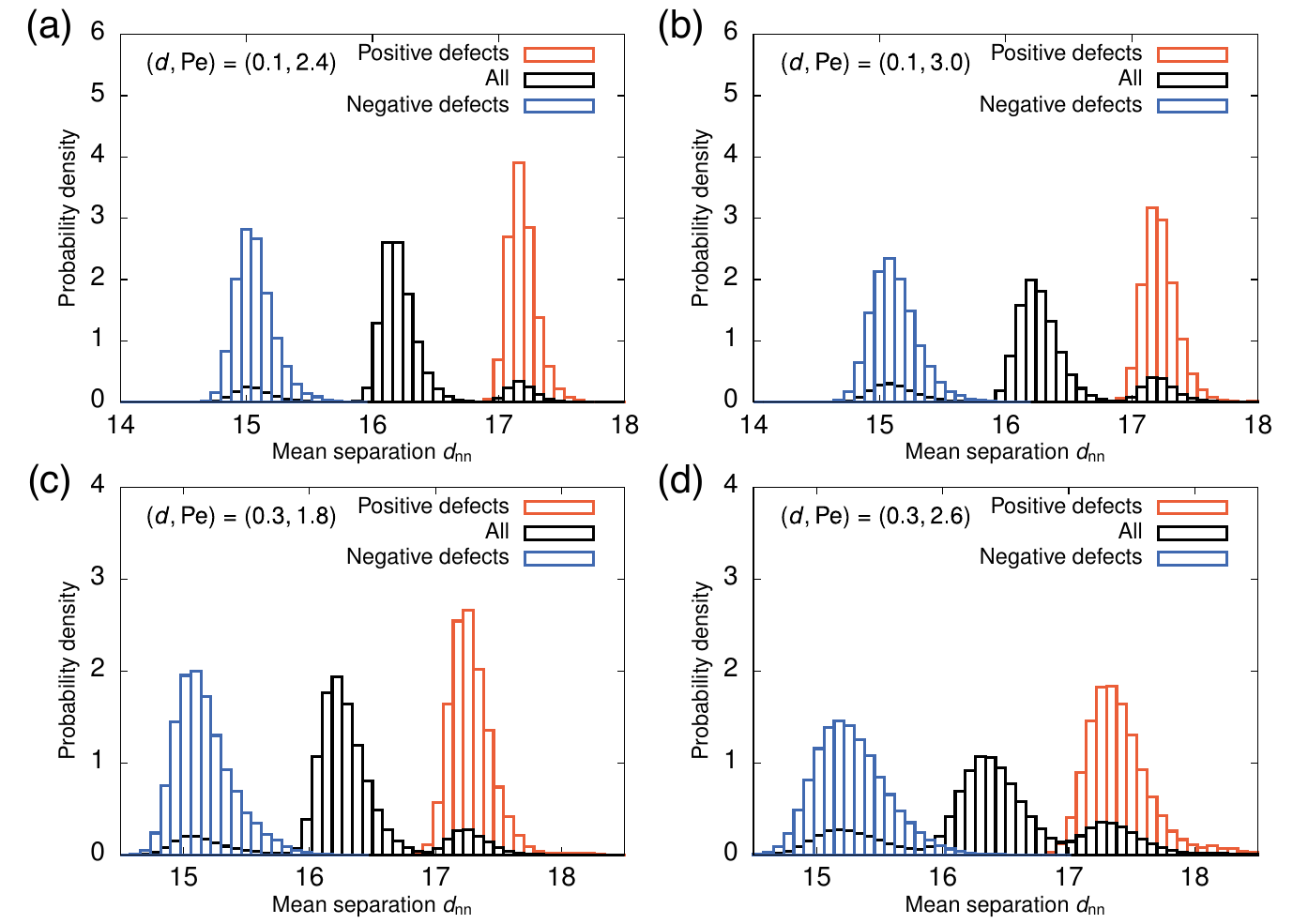}
  \mc{\caption{Probability density of the mean separation $d_{\text{nn}}$ of a cell from its nearest neighbors for negative defects, positive defects, and all cells for the same parameter sets as in Fig.~S\ref{fig:overlap_distribution}. Note that negative defects tend to be closer to their nearest neighbors than positive defects to their respective ones.}
  \label{fig:neighsep_distribution}}
\end{figure*}

\mc{We provide here a geometric explanation for the connection between defects and overlap. From measuring the contour area $A$ of each cell (Fig.~S\ref{fig:area_distribution}) and its mean separation $d_{\text{nn}}$ from its nearest neighbors (Fig.~S\ref{fig:neighsep_distribution}), we can make two observations. First, all cells have similar areas; the relative difference in the average area between negative and positive defects is around $1\%$. This is expected since we impose a severe volume constraint ($\lambda \geq 3000\,K$). Second, the average distances of cells from their nearest neighbors vary substantially depending on the defect types -- negative defects are closer to their neighbors than positive defects to their respective ones. The relative difference in the average $d_{\text{nn}}$ between negative and positive defects is about $12\%$. Indeed, the Voronoi deconstruction of the system corroborates this observation; polygons associated with negative defects have smaller areas than those associated with positive defects (see, e.g., snapshots in Fig.~3 and Fig.~S\ref{fig:overlap_defect_hexatic}). It is the combination of these two results which gives rise to a higher degree of overlap at negative defects than at positive ones.}

\begin{table*}[bht]
  \centering
  \begin{tabular}{|>{\centering}p{2cm}|>{\centering}p{4cm}|>{\centering}p{4cm}|p{5cm}<{\centering}|}\hline
    $(d,\text{Pe})$ & All vs. Positive Defects & All vs. Negative Defects & Positive vs. Negative Defects \\ \hline
    $(0.1,2.4)$ & $D = 0.31$, $p < 10^{-4}$ & $D = 0.48$, $p < 10^{-4}$ &
                  $D = 0.71$, $p < 10^{-4}$ \\
    $(0.1,3.0)$ & $D = 0.27$, $p < 10^{-4}$ & $D = 0.39$, $p < 10^{-4}$ &
                  $D = 0.63$, $p < 10^{-4}$ \\
    $(0.3,1.8)$ & $D = 0.034$, $p < 10^{-4}$ & $D = 0.20$, $p < 10^{-4}$ &
                  $D = 0.21$, $p < 10^{-4}$ \\
    $(0.3,2.6)$ & $D = 0.089$, $p < 10^{-4}$ & $D = 0.11$, $p < 10^{-4}$  &
                  $D = 0.19$, $p < 10^{-4}$\\ \hline
  \end{tabular}
  \caption{Kolmogorov-Smirnov (KS) two-sample test results for determining the significance in the deviation between the distributions of the local overlap values for negative defects, positive defects, and all cells. A higher $D$ value suggests that the two populations are more deviated from one another.}
  \label{table:overlap_ks_test}
\end{table*}

\section{SIMULATION MOVIES}
The following are the captions for the supplemental movies:
\begin{itemize}
\item Supplemental Movies 1-4: Example simulation runs showing the dynamics of the system both in the overlapping and non-overlapping regimes and both at low and high motility. The movie begins at the point when motility has just been switched on. The parameters are (Movie 1) $(d,\text{Pe})=(0.1,1.0)$, (Movie 2) $(0.1,3.0)$, (Movie 3) $(2.0,0.6)$, and (Movie 4) $(2.0,2.0)$. Same as Fig.~1 in the main paper, the coloring corresponds to the cell-index at $t = 0$.
\item Supplemental Movie 5: A simulation run showing the intermittent behavior of the system at low deformability, where it experiences episodes of freezing and melting events. The parameters are $(d,\text{Pe})=(0.1,2.4)$, and the coloring scheme is the same as above. Time series of the global bond-orientational order $\abs{\Psi_6}$ and the fraction of cells with defects $\avg{\abs{\Delta N_{\text{nn}}}}$ for this simulation run are shown at the bottom.
\end{itemize}

